\documentclass[10pt, sigconf]{acmart}

\usepackage{booktabs} % For formal tables
\usepackage{subcaption}
\usepackage{amsmath}
\usepackage{mathrsfs}
\usepackage{diagbox}
\usepackage{multirow}
\usepackage{bbm}
\usepackage[linesnumbered,ruled]{algorithm2e}
\usepackage[labelfont=bf,textfont=bf]{caption}

 \setcopyright{none}
\begin{document}
% \title{SparseMobi: Individual Vehicle Mobility Prediction with Sparse Observations}
\title{VeMo: Enabling Transparent Vehicular Mobility Modeling at Individual Levels with Full Penetration}

\author{Yu Yang}
\orcid{0000-0003-1627-5503}
\affiliation{
  \institution{Rutgers Unversity}
}
\email{yy388@cs.rutgers.edu}

\author{Xiaoyang Xie}
\affiliation{%
  \institution{Rutgers Unversity}
}
\email{xx88@cs.rutgers.edu}

\author{Zhihan Fang}
\affiliation{
  \institution{Rutgers Unversity}
}
\email{zhihan.fang@cs.rutgers.edu}

\author{Fan Zhang}
\affiliation{
  \institution{Shenzhen Institute of Advanced Technology}
}
\email{zhangfan@siat.ac.cn}

\author{Yang Wang}
\affiliation{
  \institution{University of Science and Technology of China}
}
\email{Angyan@ustc.edu.cn}

\author{Desheng Zhang}
\affiliation{
  \institution{Rutgers Unversity}
}
\email{desheng.zhang@cs.rutgers.edu}

\titlenote{Prof. D. Zhang and Y. Wang are the joint corresponding authors.}

% This work is a pre-print version to appear at MobiCom 2019.

\renewcommand{\shortauthors}{Y. Yang et al.}
% \maketitle

\begin{abstract}
Understanding and predicting real-time vehicle mobility patterns on highways are essential to address traffic congestion and respond to the emergency. However, almost all existing works (e.g., based on cellphones, onboard devices, or traffic cameras) suffer from high costs, low penetration rates, or only aggregate results.
To address these drawbacks, we utilize Electric Toll Collection systems (ETC) as a large-scale sensor network and design a system called VeMo to transparently model and predict vehicle mobility at the individual level with a full penetration rate.
Our novelty is how we address uncertainty issues (i.e., unknown routes and speeds) due to sparse implicit ETC data based on a key data-driven insight, i.e., individual driving behaviors are strongly correlated with crowds of drivers under certain spatiotemporal contexts and can be predicted by combining both personal habits and context information.
More importantly, we evaluate VeMo with 
(i) a large-scale ETC system with tracking devices at 773 highway entrances and exits capturing more than 2 million vehicles every day;
(ii) a fleet consisting of 114 thousand vehicles with GPS data as ground truth.
We compared VeMo with state-of-the-art benchmark mobility models, and the experimental results show that VeMo outperforms them by average 10\% in terms of accuracy.
\end{abstract}




\maketitle

\section{Introduction}

Understanding and modeling individual vehicular mobility on highways have various applications, e.g., congestion prediction~\cite{huang2010dynamic}, 
route planning~\cite{ben2010road} and ramp metering~\cite{schmidt2015decentralised}.
However, modeling and predicting individual vehicle locations in fine spatial-temporal granularity are extremely challenging due to a large number of vehicles and limited infrastructures on highways compared to cities~\cite{511nj}\cite{nyc_camera}.

The existing approaches for vehicle location prediction can be basically categorized into two groups:
(i) mobile infrastructure based solutions such as cellphones (e.g., Online Map Services~\cite{google_map}) and onboard devices (e.g., OBD devices~\cite{onstar}), and
(ii) static infrastructure based solutions: traffic cameras~\cite{wan2014camera}, loop sensors~\cite{taghvaeeyan2014portable}, and RFID~\cite{yang2014tagoram}.
For mobile infrastructure based solutions, they typically have privacy issues since they require real-time GPS locations of vehicles~\cite{zang2011anonymization};
for static infrastructure based solutions, they typically introduce low spatial coverage or high costs for a complete highway system coverage~\cite{itscosts}.
Further, both of them may suffer low penetration rates, e.g., some commuters do not use navigation apps when traveling some familiar routes~\cite{statista}; traffic cameras are not pervasive on highways in some countries~\cite{speed_camera}.

In this paper, to address these drawbacks, we utilize a highway Electric Toll Collection (ETC) system as a sensor network for vehicular mobility modeling and prediction.
Compared to the existing approaches, our ETC based solution has the following features:
(i) it requires no additional infrastructure since it relies on data already gathered in real time over highway networks for toll collections;
(ii) it poses no additional privacy threats because it does not collect vehicle-specific GPS data;
(iii) it does not suffer from low penetration rates since all vehicles have to be charged by an ETC system when using highway systems.
% MODIFY
Even some highways are installed with induction loops, they cannot achieve individual level modeling compared to the ETC system.

However, since an ETC system is deployed for toll collections instead of mobility modeling, we have the following new challenges.
(i) An ETC system only logs when and where a vehicle enters and leaves a highway system for billing purposes and it leads to extremely sparse location records for each vehicle, i.e., only two data points per trip, which makes predicting destinations without intermediate locations be challenging. Without any historical routes or speeds logged, it is difficult to train a model.
(ii) In a complicated highway network, given an entrance and exit, there are many potential routes as shown by our later analyses, and ETC data do not log any information regarding which route was taken during a particular origin and destination pair. Based on our data, we found that the shortest routes are not the first choices for many vehicles due to congestion.
(iii) Traffic speeds vary by different spatiotemporal contexts, and ETC data do not directly log speeds. Straightforward solutions (e.g.,  assuming real-time speeds vary near the speed limit) usually do not perform well because of various driving behaviors under different contexts.

To address these challenges, in this paper, we perform a systemic investigation of a large-scale ETC system along with its data, and we found a key data-driven insight:  
even with complicated highway networks and real-time context, individual travel behaviors are strongly correlated with crowds under certain spatiotemporal contexts and can be predicted by combining both personal habits and context information.
% individual travel behaviors (in terms of origins, destinations, entry times, exit times, routes) are regular under certain spatiotemporal contexts at both an individual level and group level.
Built upon this insight, we design a model called VeMo to model and predict individual vehicular mobility patterns based on sparse observations on real-time origins as well as historical origins and destinations only. 
In particular, the key contributions of this paper are as follows.

\begin{itemize}

\item

To our knowledge, we conduct the first systematic investigation of real-time vehicular mobility modeling and prediction based on large-scale ETC and GPS data.
Our investigation is based on real-time and historical ETC data from 7.8 million vehicles and GPS data from 114 thousand vehicles.
This large-scale vehicle sensing study enables us to find mobility insights that are not possible to obtain with small-scale systems and data. 
By working with our collaborators, we released some processed sample data for the benefit of the research community\footnote{https://www.cs.rutgers.edu/~dz220/}.

\item

We analyze both ETC and GPS data and provide some in-depth discussions on vehicular mobility patterns on highways.
Based on the insights from our analyses, we design a mobility prediction system called VeMo with three key components to predict destinations, routes, and speeds for individual vehicles based on both historical and real-time ETC data.
Technically, we extract unobserved routes and speeds through a joint optimization model. By studying various mobility features at both the individual level and crowd level, we fuse them based on a Mondrian Forests model to address the uncertainty issue in the mobility prediction.
% Another key feature of VeMo is in its self-awareness with which based on real-time events (e.g., a vehicle was predicted to exit but it did not, or a vehicle exited earlier than predicted), it will automatically adjust prediction models for better performance.

\item More importantly, we implement and evaluate the VeMo in Guangdong Province, China with
(i) an ETC system covering 1,439 highway entrances and exits, and it captures around 2 million vehicles per day;
(ii) a vehicle fleet and its GPS data including 114 thousand vehicles for evaluation only, where 20\% of vehicles have the trajectories on highways.

\item We evaluate VeMo through a two-month set of ETC and GPS data by showing both intermediate results (e.g., predicting destinations, routes, and speeds) and end-to-end results (e.g., predicting real-time locations).
We study the performance sensitivity of our system to different spatial-temporal contexts. Compared with state-of-the-art solutions, VeMo provides a $10\%$ performance gain on average in terms of prediction accuracy.

\end{itemize}

% \noindent The rest of the paper is organized as follows.
% Section 2 presents our motivation.
% Section 3 introduces our ETC system. 
% Section 4 provides the detailed design for VeMo.
% Section 5 evaluates the performance of VeMo.
% Section 6 discusses insights and limitations,
% followed by the related work and conclusion in Sections 7 and 8.
\section{Motivation}
\label{sec:moti}

\subsection{Use cases}
VeMo aims to predict the real-time locations of individual vehicles, which enables various applications that cannot be achieved by previous solutions. As collaboration with the highway administrators, we gives two exemplary applications that matter a lot to the highway management.

\begin{itemize}

\item \textbf{Highway anomaly detection:} One important task for highway administrators is to detect the highway anomaly at the first time, such as traffic accidents. However, it is quiet expensive to arrange regular road check manually or cannot detect anomalies in time. Through predicting the real-time location of a vehicle, we can know when the vehicle is expected to leave the highway in the regular situation. Conversely, we could know there may be an anomaly event if a number of vehicles do not leave the highway as expected.

\item \textbf{Highway risk assessment:}
Improving driving safety on highways is always an important topic for the highway administration companies. Noticeably, there are more than 6 million crashes on highways in the United States during 2015, including more than 30 thousand fatalities and 1 million injuries ~\cite{highway_fatal}.
By transparently predicting the locations of individual vehicles, highway administration companies can understand the number of affected vehicles if there were an accident on certain road segments, and provide some contingency plans accordingly. Another safety related application is to localize a vehicle of interest (e.g., a vehicle with dangerous cargo or suspects) for public safety after it enters the highway.

% \item \textbf{Adaptive Dynamic Tolls:}
% Dynamic tolls have received attentions for a long time to control traffic congestion on highways, especially from highway administration companies~\cite{wired}. However, the current strategies~\cite{phan2016model} are only designed according to limited historical information (e.g., the number of vehicles entering the highways from certain toll stations), which only provides partial observations of the highway traffic rather than the real-time fine-grained throughput on each road segment. To obtain fine-grained throughput, one typical solution is to deploy sensors (e.g., loop sensors) on all the road segments, which is quite expensive~\cite{loop_cost}. Instead, our solution can be applied to estimate throughput on each road segment in real time through predicting the real-time location of each vehicle, which enables more sophisticated strategies with adaptation on the real-time traffic conditions.
\end{itemize} 

\noindent\textbf{Uniqueness of ETC based systems:}
To implement those applications, previous works either require extra installed infrastructures or suffer from low penetration rates of vehicles.
For example, mobile phone based solutions can only know the locations of a number of vehicles, which cannot provides the accurate number of vehicles in a certain location.
Induction loop based solutions cannot identify the uniqueness of a vehicle.
Traffic cameras are potentially used to detect individual vehicles but limited by the laws in many countries such as U.S.
Moreover, in developing countries, satellite images or mobile infrastructure is not well penetrated and it is really hard to predict the real-time locations. The ETC based toll system is universal and exist almost everywhere even in developing countries. 
Therefore, ETC based systems utilize widely deployed infrastructure (i.e., ETC), which can transparently obtain information from vehicles (i.e., when charging toll) with extremely low marginal cost. Moreover, the full penetration rate on highways can also make up for the weakness of mobile phone based solutions.

% Other infrastructures such as speed cameras are not pervasive on highways in some countries~\cite{speed_camera}. The detailed discussion is given in Section~\ref{sec:soa}.

% \noindent\textbf{(ii) Traffic anomaly detection:}
% When detecting the traffic anomaly, current solutions are mainly based on crowdsourcing~\cite{waze}, which are suffered from either the low participation of drivers or unsafe driving behaviors or low quality~\cite{app_accidents} of crowdsourcing~\cite{wazefool}. Our system presents a self-aware solution to predict the real-time locations of vehicles, which only relies on the existing infrastructure and relaxes the participation of people. Through analyzing the abnormal mobility pattern of crowd vehicles (i.e., abnormal delay), we can efficiently detect the possible traffic anomalies.

\subsection{Challenges}

It is not trivial to predict the real-time locations of vehicles because of the uncertainties caused by various traffic conditions and driving behaviors. To show these challenges, we study one-month data (both ETC transactions and trajectories of sample vehicles) in the Guangdong province of China and identify several challenges regarding three key factors including destinations, routes and speeds. The detailed data description is presented in Section~\ref{sec:infras} and Section~\ref{sec:eval}.

\begin{figure}[htbp] \centering
    % \vspace*{-5pt}
    \begin{minipage}{0.5\linewidth} \centering
        % \vspace*{-5pt}
%         \hspace*{-30pt}
        \includegraphics[width=\linewidth, keepaspectratio=true]{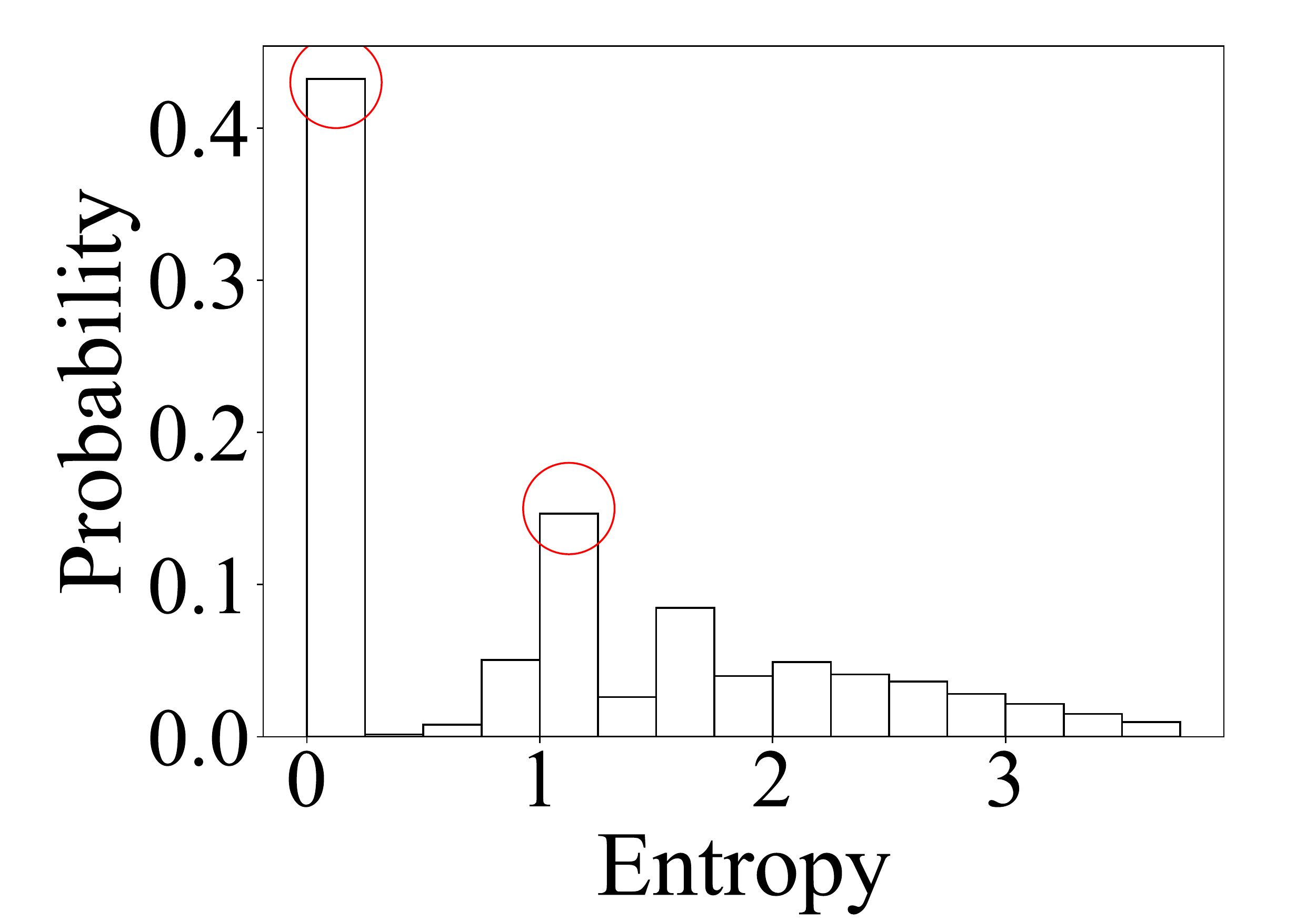}
        % \vspace*{-20pt}
        \caption{Destination entropy}
        \label{fig:dest_entropy}
        % \hspace*{5pt}
    \end{minipage}
    \hspace*{-5pt}
    \begin{minipage}{0.5\linewidth} \centering
        % \vspace*{-5pt}
%         \hspace*{-30pt}
        \includegraphics[width=\linewidth, keepaspectratio=true]{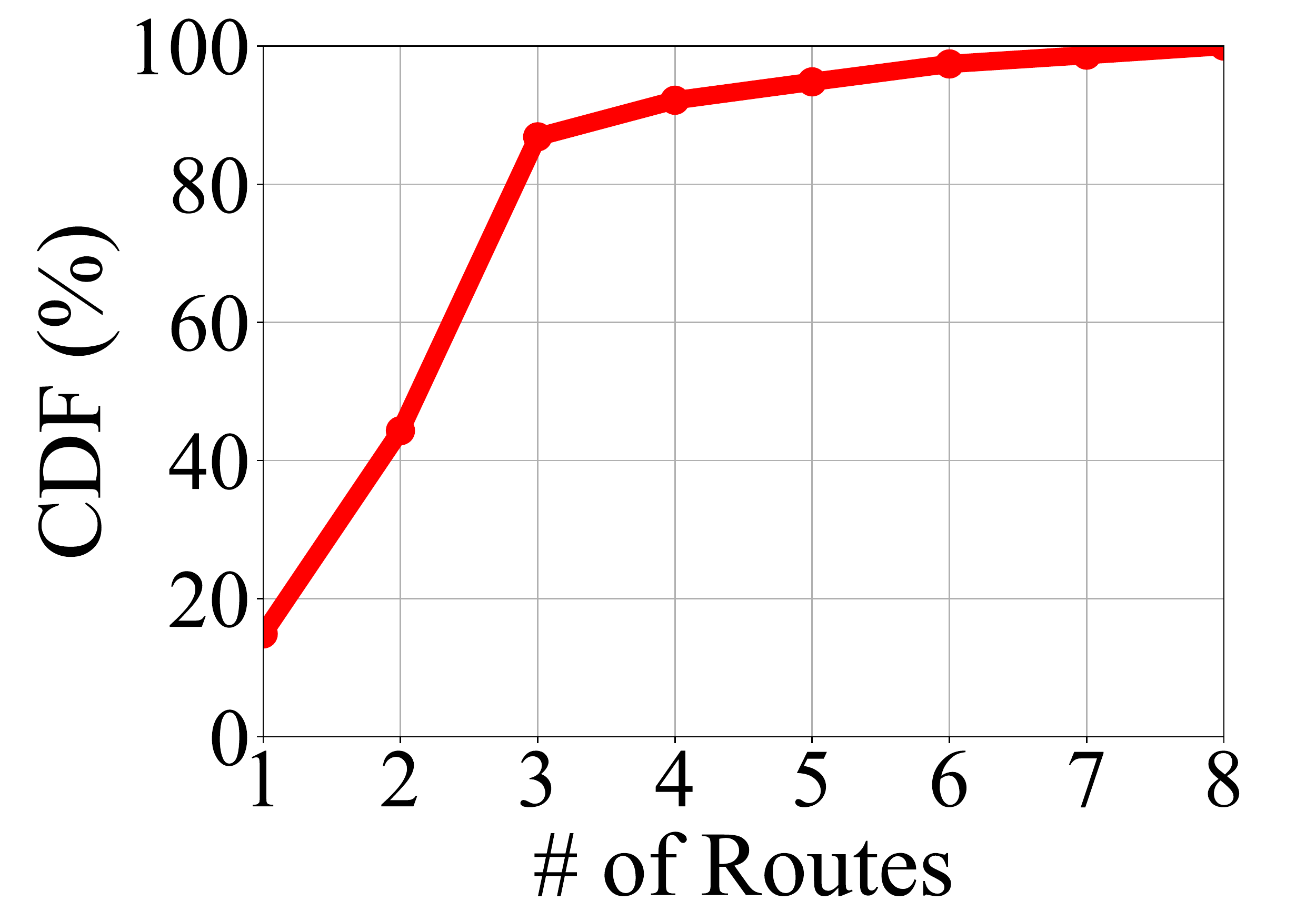}
        % \vspace*{-20pt}
        \caption{Number of routes}
        \label{fig:num_routes}
    \end{minipage}
    % \vspace*{-5pt}
\end{figure}

\noindent\textbf{(i) Destination uncertainty:}
To predict the real-time locations of vehicles, it is important to understand the destinations and routes. However, it is not trivial to predict the routes and destinations. To characterize the inherent predictability across vehicles, we present the destination entropy of each vehicle in Fig~\ref{fig:dest_entropy}. The figure reveals two peaks as the entropy equals 0 and 1, which indicates the next location of a vehicle could be found on average in any $2^0=1$ and $2^1=2$ locations, respectively. Especially, we find most vehicles travel on highways only once in one month when the entropy=0; vehicles are more like to commute between two locations when the entropy=1.
Many works~\cite{xu2016destpre}~\cite{dewri2013inferring} have been done to predict the destinations of vehicles whose entropy is greater than or equal to one since those vehicles generally have regular commute patterns or extensive historical data. 
However, it is not clear how to predict the destinations of vehicles with only a few historical transactions. We refer this problem as a \textit{destination sparsity} problem.

\noindent\textbf{(ii) Unobserved routes and speeds:} Previous studies have been done to model the route choices and driving speeds~\cite{ben2010road}~\cite{yu2016senspeed}. 
Through studying the historical routes and speeds in the trip recorded by GPS-based devices, some sophisticated models are proposed to predict vehicular mobility in the near future. 
However, in our setting, one of the key characteristics of the ETC system is that it can only obtain very sparse information (i.e., the time and location when entering and exiting highways). 
This leads to the problem that we cannot obtain detailed routes and speeds to learn the route choice model and the driving speed model, which is not solved in the previous work.

\begin{figure}[htbp] \centering
    % \vspace*{-5pt}
    \begin{minipage}{0.43\textwidth} \centering8
        % \vspace*{-5pt}
        \includegraphics[width=\textwidth, keepaspectratio=true]{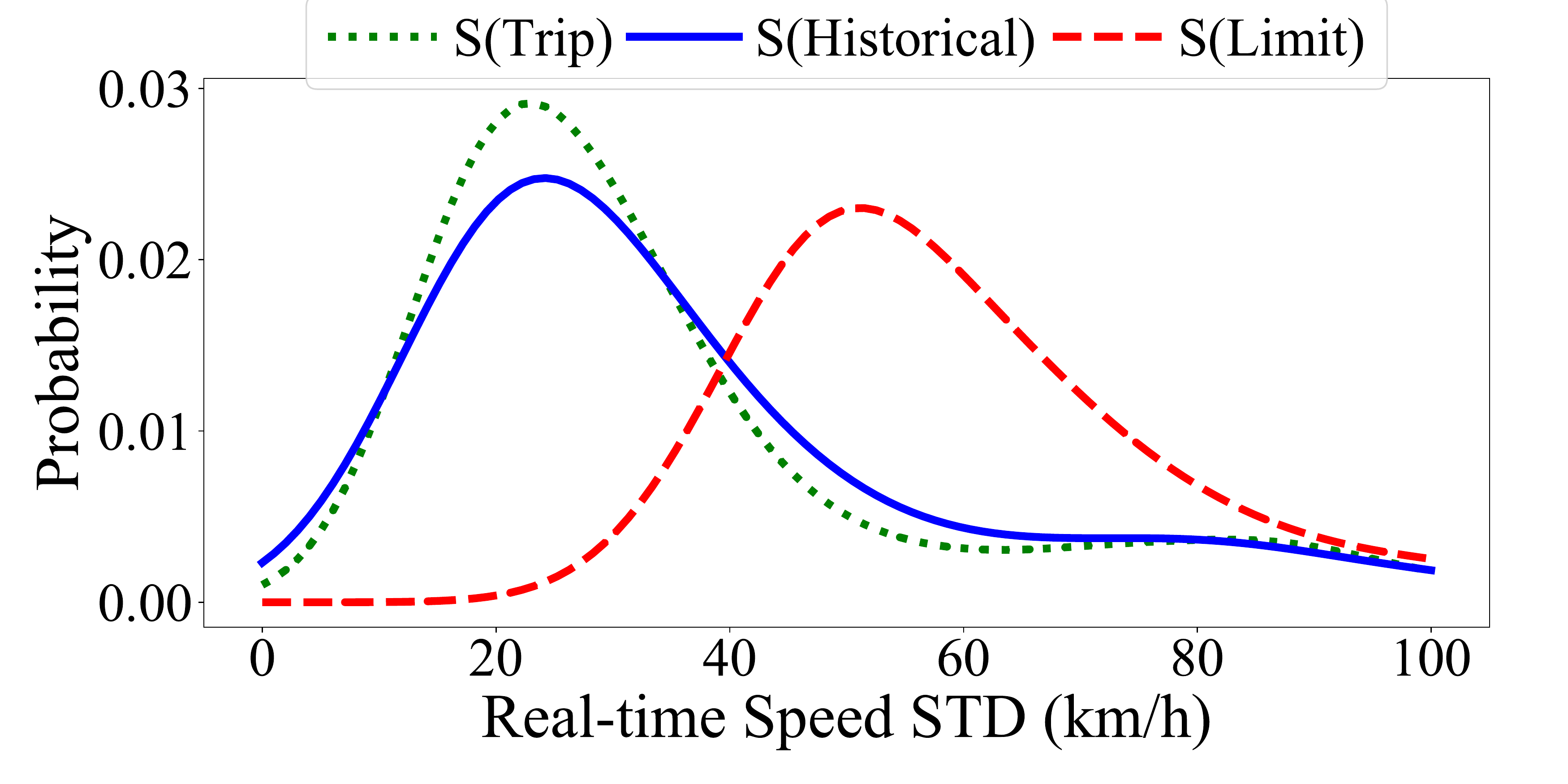}
        % \vspace*{-20pt}
    \end{minipage}
    \caption{Speed STD}
    \label{fig:speed_std}
    % \vspace*{-5pt}
\end{figure}
Moreover, routes and speeds also vary depending on user behaviors and contexts. 
For a given origin and destination, people can choose different routes if the road network is not trivial (i.e., only one route from the origin to the destination). Fig~\ref{fig:num_routes} illustrates the number of routes between the origin-destination pairs.
We found that only $17\%$ of station pairs have only one route based on GPS trajectories obtained from 114 thousand vehicles.
It is impractical to assume only shortest routes are used by vehicles.
(Note that these trajectories are only used in the motivation and evaluation rather than the model design.)
As for speeds, people empirically expect that the driving speeds of vehicles are around certain speeds (e.g., speed limit or average speed) with less variance.
However, in our study, we found the real-time speed is more complicated than the empirical intuition. 
To illustrate the characteristics of real-time speed, we study the real-time speed standard deviation (STD) across vehicles by replacing the mean value in the standard formula of standard deviation with the speed limit($S(Limit)$), the historical average speed($S(Historical)$), the current trip average speed($S(Trip)$), respectively.
Fig~\ref{fig:speed_std} demonstrates both $S(Historical)$ and $S(Trip)$ have a Gaussian-like distribution with the mean STD near 20 $km/h$.
It leads to a 330-meter offset in one-minute driving if only the average speed is utilized to obtain the real-time location.
It also revels the fact that it is difficult for people to drive at the speed limit (e.g., can only drive at 60 km/h compared to the speed limit of 120 km/h) because of the heavy traffic.

\subsection{Summary} The ETC based system provides an unprecedented opportunity to transparently model and predict vehicular mobility with a full penetration rate, which enables various potential applications such as highway safety management and adaptive dynamic toll strategies. However, due to the unique characteristic of only observing vehicles at entrances and exits, there are several challenges to be solved including destination sparsity problem and unobserved routes and speeds.

\section{ETC System and Data Description} 
\label{sec:infras}
We first introduce some notations to facilitate our discussion, and then give a brief description of an ETC system based on our infrastructure access in Guangdong and finally provide some data-driven insights.

\noindent\textbf{Notations:} Given ETC data on the vehicle's trip levels, 

\begin{itemize}
    \item An \textbf{edge} $e$ is a highway segment between two adjacent toll stations, i.e., the finest spatial unit for ETC data-based modeling.

 \item A \textbf{route} $r$ is a set of adjacent edges, which connect the origin toll station and the destination toll station of a particular trip. 

 \item A \textbf{K-edge trip} is a trip of a particular vehicle with $K$ edges in its route between the origin and the destination. 
 Specifically, a \textbf{single-edge trip} has only one edge in the route.
\end{itemize}

\noindent Based on the above terms, our problem definition is 
"Given a vehicle entering a highway network from a toll station as an origin $S_o$ at time $T_o$, predict its real-time locations on highways at any given time $T_r$ until it exits the highway.

\begin{figure}[htbp] \centering
    \begin{minipage}{0.48\textwidth} \centering
        % \vspace*{-5pt}
        \includegraphics[width=\textwidth, keepaspectratio=true]{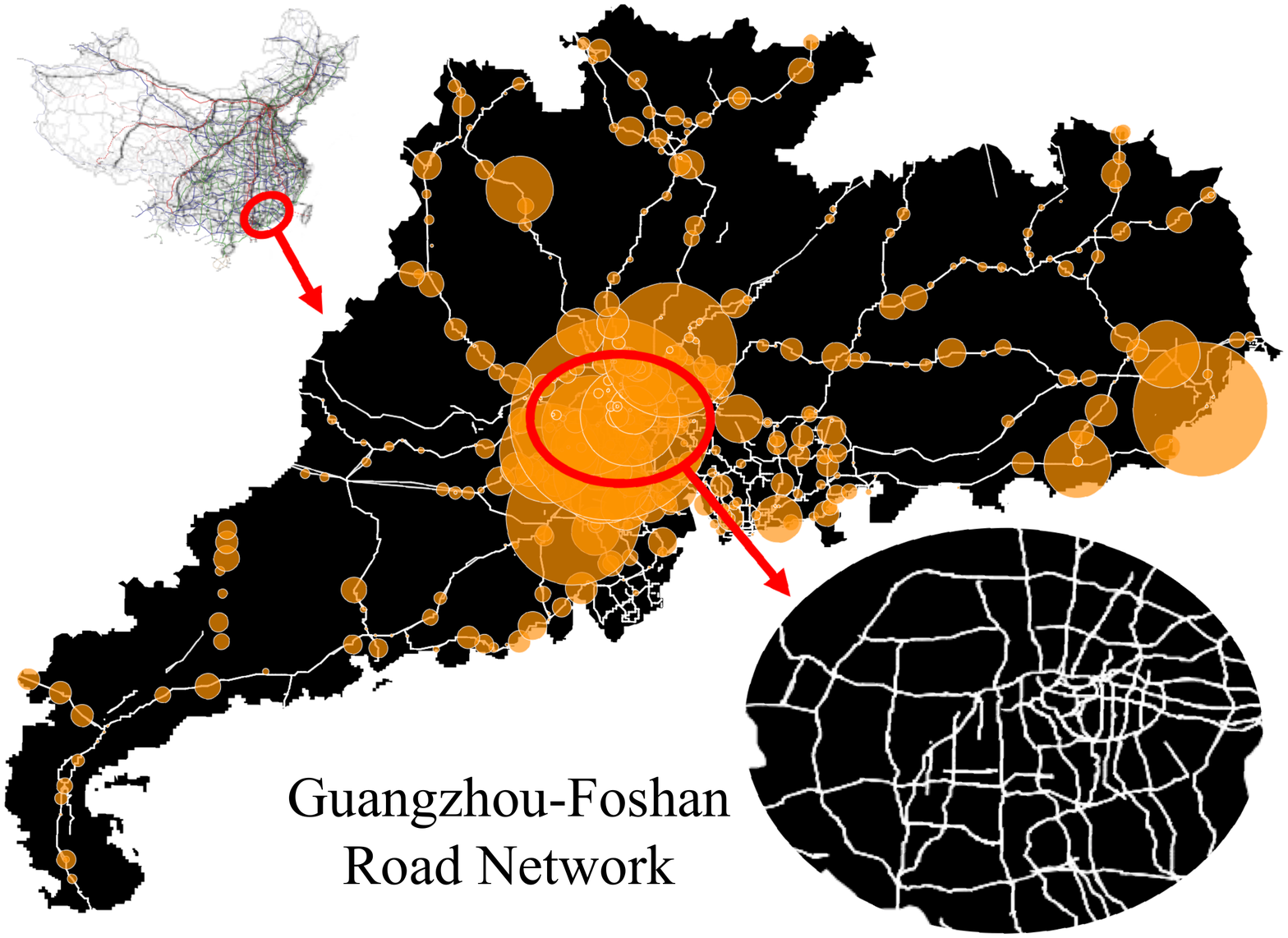}
        % \vspace*{-20pt}
    \end{minipage}
    \caption{ETC Systems in Guangdong Province}
    \label{fig:etc}
    % \vspace*{-5pt}
\end{figure}

\noindent\textbf{Infrastructure Overview:}
Fig~\ref{fig:etc} shows the road structure and the locations of toll stations in the Guangdong province, which has 69 highways and 773 ETC toll stations with 1,439 highway entrances and exits covering an area of $179,800km^2$.
The circles represent toll stations and the larger the icon, the heavier the daily traffic volume.
It shows the traffic mainly concentrates on the central area and the road structure in that area is also complex as shown in the Guangzhou-Foshan Road Network.
Each toll station detects all vehicles when they enter the highway system, and then logs the records as transactions after they leave the highway system.
The toll station identifies a vehicle by ETC RFID devices (for regular charging) or cameras (for the purpose of detecting escaping charges).

\begin{table}[htbp]
\centering
\caption{ETC Transaction Description}
\label{tb:etc_data}
% \vspace*{-10pt}
\begin{tabular}{@{}ll@{}}
\toprule
Field                      & Value               \\ \midrule
Entering/Exit Toll Station & Humen Station       \\
Entering/Exit Time         & 2016-07-01 13:00:01 \\
Vehicle Id                 & F37SS1D4GU          \\
Vehicle Type               & Car/Bus/Truck       \\
Axis Count                 & 2                   \\
Weight                     & 1500kg              \\ \midrule
\multicolumn{2}{l}{Number of Daily Transactions: 4 millions} \\ \multicolumn{2}{l}{Number of Daily Vehicles: 2 millions} \\ \bottomrule
\end{tabular}
\end{table}

As shown in Table~\ref{tb:etc_data}, each generated transaction contains information including entering and exit station, entering and exiting time, vehicle id, vehicle type (i.e., car, bus, truck), axis count and weight. Such a transaction was generated when a vehicle enters and exits the highway network with both ETC cards or cash.
On average, there are more than 4 million transactions generated every day from 2 million vehicles.

\begin{figure}[htbp] \centering
    \vspace*{5pt}
    \begin{minipage}{0.5\linewidth} \centering
%         \hspace*{-30pt}
        \includegraphics[width=\linewidth, keepaspectratio=true]{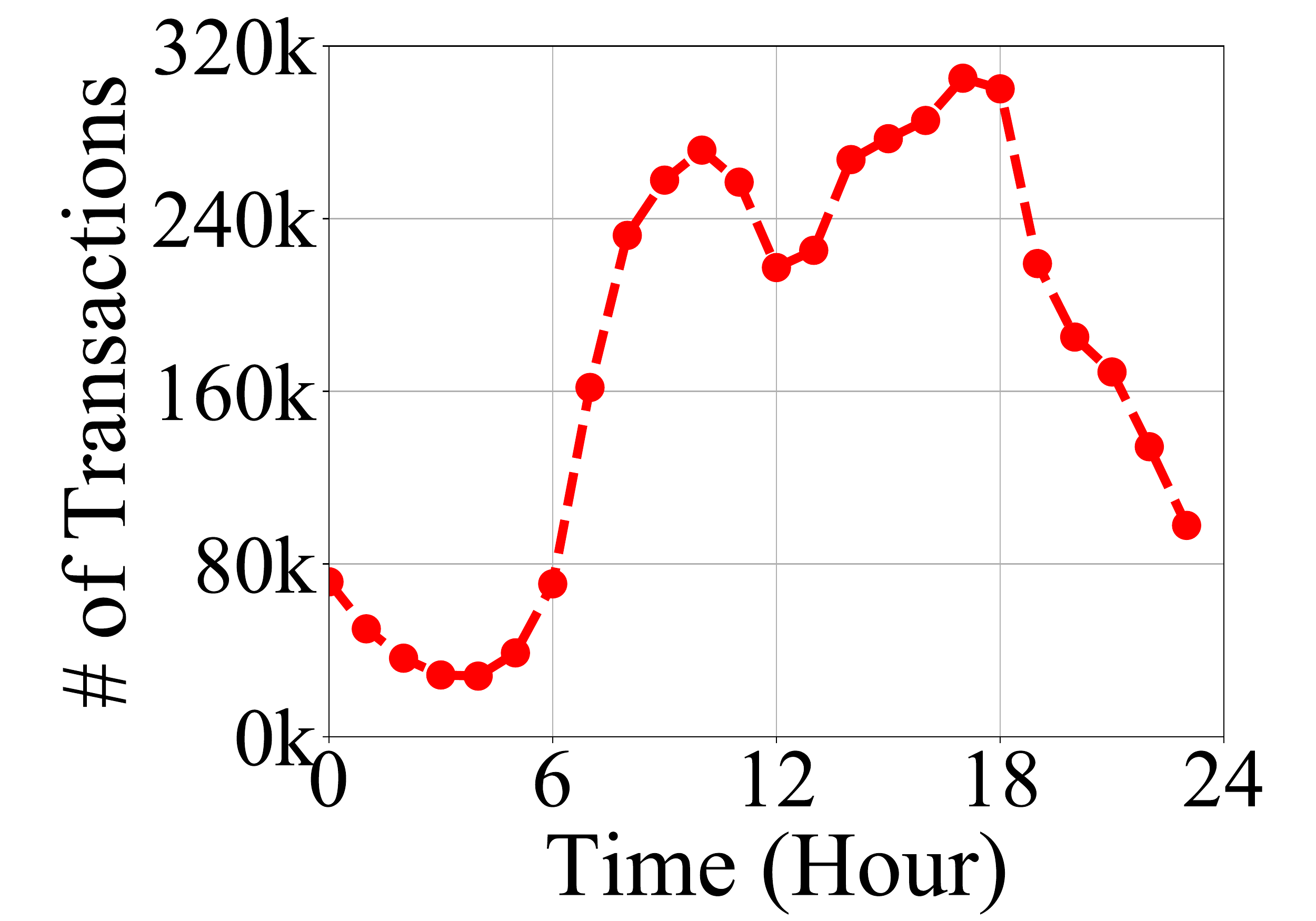}
        % \vspace*{-15pt}
        \caption{Volume over time}
        \label{fig:volume_time}
        % \hspace*{5pt}
    \end{minipage}
    \hspace*{-5pt}
    \begin{minipage}{0.5\linewidth} \centering
%         \hspace*{-30pt}
        \includegraphics[width=\linewidth, keepaspectratio=true]{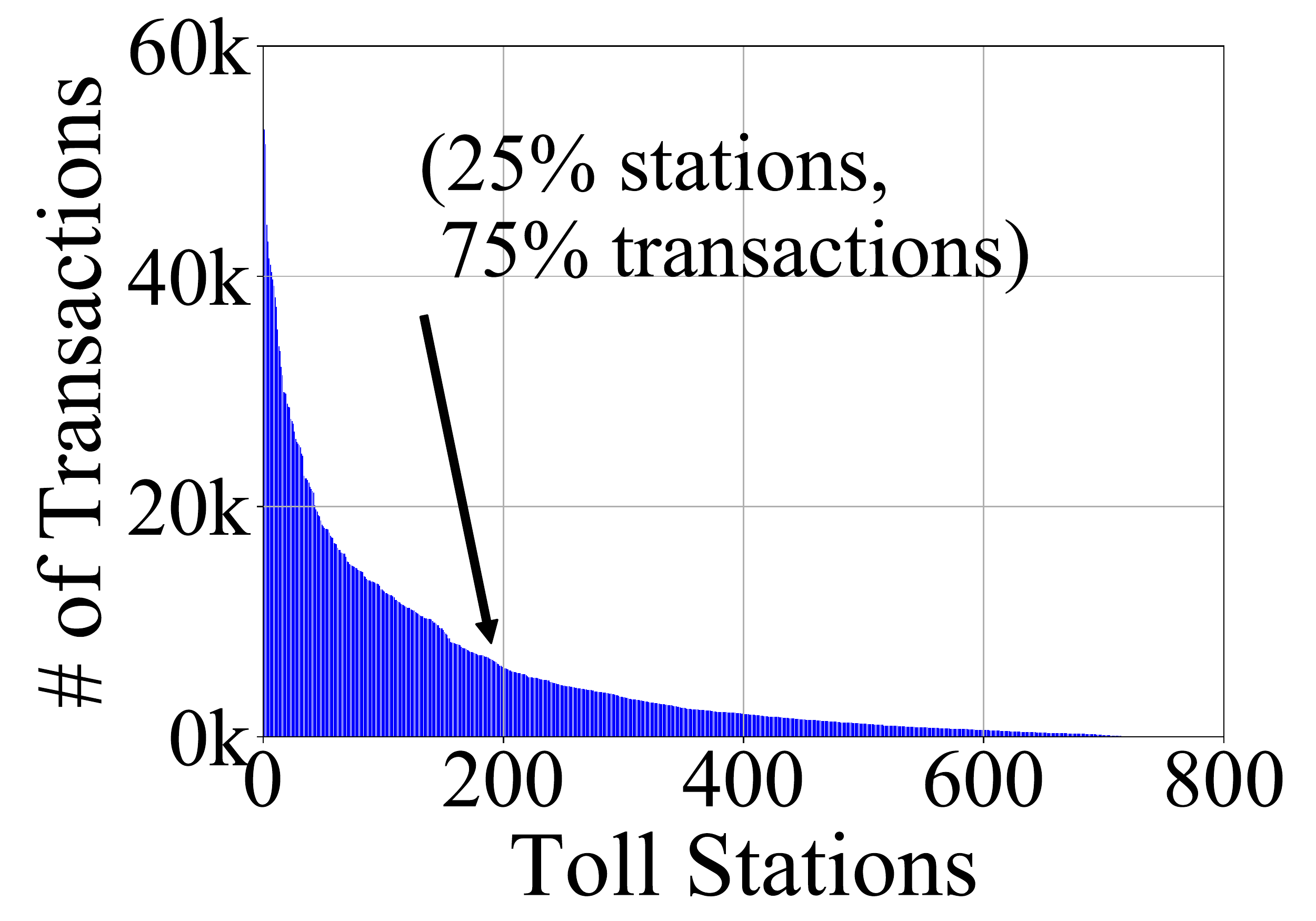}
        % \vspace*{-15pt}
        \caption{Volume over stations}
        \label{fig:volume_pdf}
    \end{minipage}
    % \vspace*{-5pt}
\end{figure}

\noindent\textbf{Statistic description:} Fig~\ref{fig:volume_time} plots the average traffic volume in 24 hours of a day. 
It shows there are two peak hours (i.e., 10 am and 6 pm), which potentially make prediction challenging due to uncertainty (e.g., route choice, traffic jam, etc) introduced by high traffic volume.
Fig~\ref{fig:volume_pdf} depicts the daily transaction volume of all the toll stations, where 25 $\%$ of the stations contribute 75 $\%$ of the transactions. 
It suggests the major number of vehicles enter the highway from a limited number of stations, indicating prediction related to unpopular stations may suffer from lack of historical and real-time data.

\begin{table*}[!htbp]
\centering
\caption{Mobility Modeling Features}
\label{tb:features}
% \vspace{-10pt}
\begin{tabular}{|l|l|l|l|}
\hline
 & \begin{tabular}[c]{@{}l@{}}Individual Features\end{tabular}                           
 & \begin{tabular}[c]{@{}l@{}}Crowd Features\end{tabular} 
 & \begin{tabular}[c]{@{}l@{}}Context Features\end{tabular}        
 \\ \hline
\begin{tabular}[c]{@{}l@{}}Destination Predictor\\(Section~\ref{sec:destination_prediction})\end{tabular} & \begin{tabular}[c]{@{}l@{}}
Historical Destinations,\\ Time of Day, Vehicle Type\end{tabular}   
& Crowd Destination Distributions                 
&\begin{tabular}[c]{@{}l@{}} Day of Week\\ Weekday/weekend\end{tabular} 
\\ \hline
\begin{tabular}[c]{@{}l@{}}Route Predictor\\(Section~\ref{sec:route_prediction})\end{tabular}             & \begin{tabular}[c]{@{}l@{}}Historical Routes (Section ~\ref{sec:historical_data}),\\ Driving Experience, Time of Day \end{tabular} & Crowd Route Distributions                                
&\begin{tabular}[c]{@{}l@{}} Day of Week\\ Traffic Speed  \end{tabular} 
\\ \hline
\begin{tabular}[c]{@{}l@{}}Speed Predictor\\(Section~\ref{sec:speed_prediction})\end{tabular}             & \begin{tabular}[c]{@{}l@{}}Historical Driving Speed (Section ~\ref{sec:historical_data}),\\ Time of Day, Vehicle Type\end{tabular}  
& Crowd Speed Distributions                                
& \begin{tabular}[c]{@{}l@{}}Weekday/weekend,\\ Weather\end{tabular} 
\\ \hline
\end{tabular}
\end{table*}

\section{VeMo Design}
\label{sec:model}

In this section, we first depict the overview framework of VeMo, which is then followed by feature extraction of three components including 
(i) destination prediction, 
(ii) route inference, and
(iii) speed estimation.
Specifically, in the route and speed inference, we utilize a joint optimization model to learn the historical routes and speeds with only transaction data, to obtain necessary training data.
These features are fed into a learner to learn predictors for different tasks.

\begin{figure}[htbp] \centering
    % \vspace*{-20pt}
    \begin{minipage}{0.48\textwidth} \centering
        \includegraphics[width=\textwidth, keepaspectratio=true]{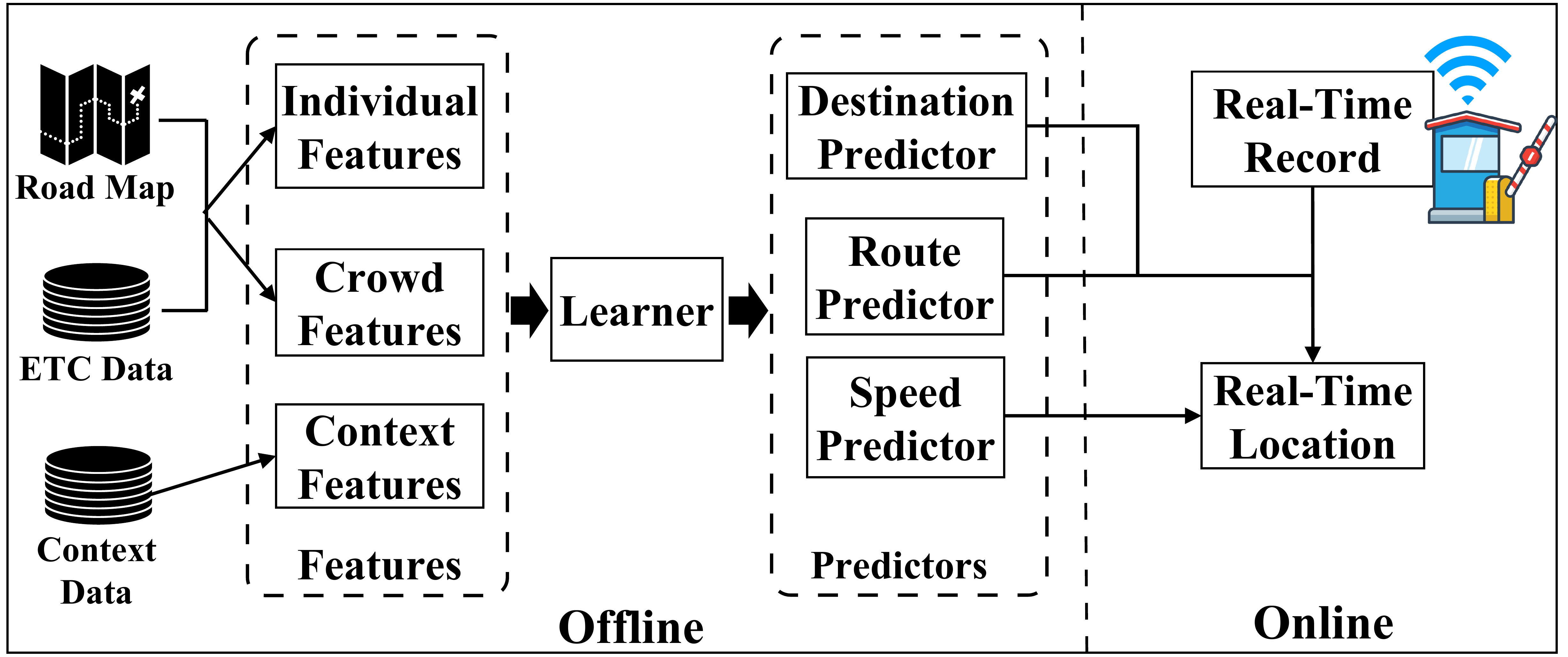}
    \end{minipage}
    \caption{Framework}
    \label{fig:model}
\end{figure}

Fig~\ref{fig:model} shows the framework of our system, which consists of two parts: offline learning and online prediction.
\noindent In the offline learning, all the data come from three data sources including the road map, historical ETC transactions and context data.
In the feature extraction, we divide all the features into three categories, which are individual features, crowd features and context features. 
The feature summary is presented in Table~\ref{tb:features} (next page).
By fitting these features into the learner, we train three predictors for destinations, routes and speeds. By combining these predictor together, we predict the real-time locations of vehicles.
In the next three subsections, we introduce three predictors for destinations, routes, and speeds from a feature perspective respectively, and then unify them together with a prediction model based on Mondrian Forest.

\subsection{Framework}

\subsection{Destination Predictor}
\label{sec:destination_prediction}

Destination prediction has been intensively studied in the past few years~\cite{gonzalez2008understanding,zhang2014exploring}. The existing approaches for the vehicle destination prediction mainly rely on transition probabilities between different locations through learning historical trajectories using various Markov chain based models~\cite{li2015inferring, do2012contextual}. One of the key prerequisites is that there should be enough historical data of individuals to learn the transition probabilities. However, in our context, most vehicles only have limited historical data (as we discussed in Section~\ref{sec:moti}), which makes it hard to directly apply the Markov chain based models. To address this issue, we explore more individual features, crowd features and context features. 

\begin{figure}[htbp] \centering
    % \vspace*{-5pt}
    \begin{minipage}{0.5\linewidth} \centering
%         \hspace*{-30pt}
        \includegraphics[width=\linewidth, keepaspectratio=true]{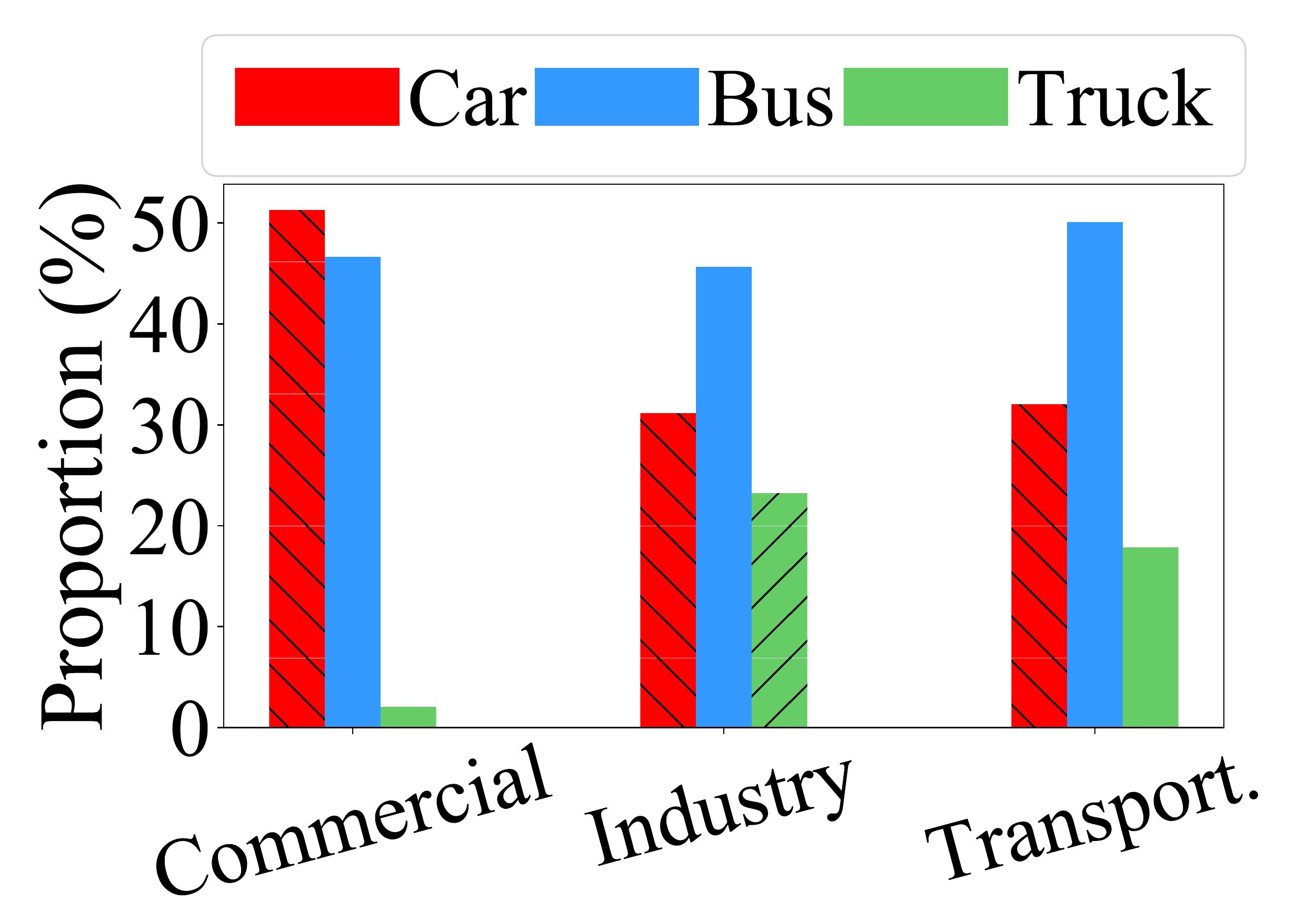}
        % \vspace*{-20pt}
        \caption{Dest. variance}
        \label{fig:v_type_dest}
        % \hspace*{5pt}
    \end{minipage}
    \hspace*{-5pt}
    \begin{minipage}{0.5\linewidth} \centering
%         \hspace*{-30pt}
        \includegraphics[width=\linewidth, keepaspectratio=true]{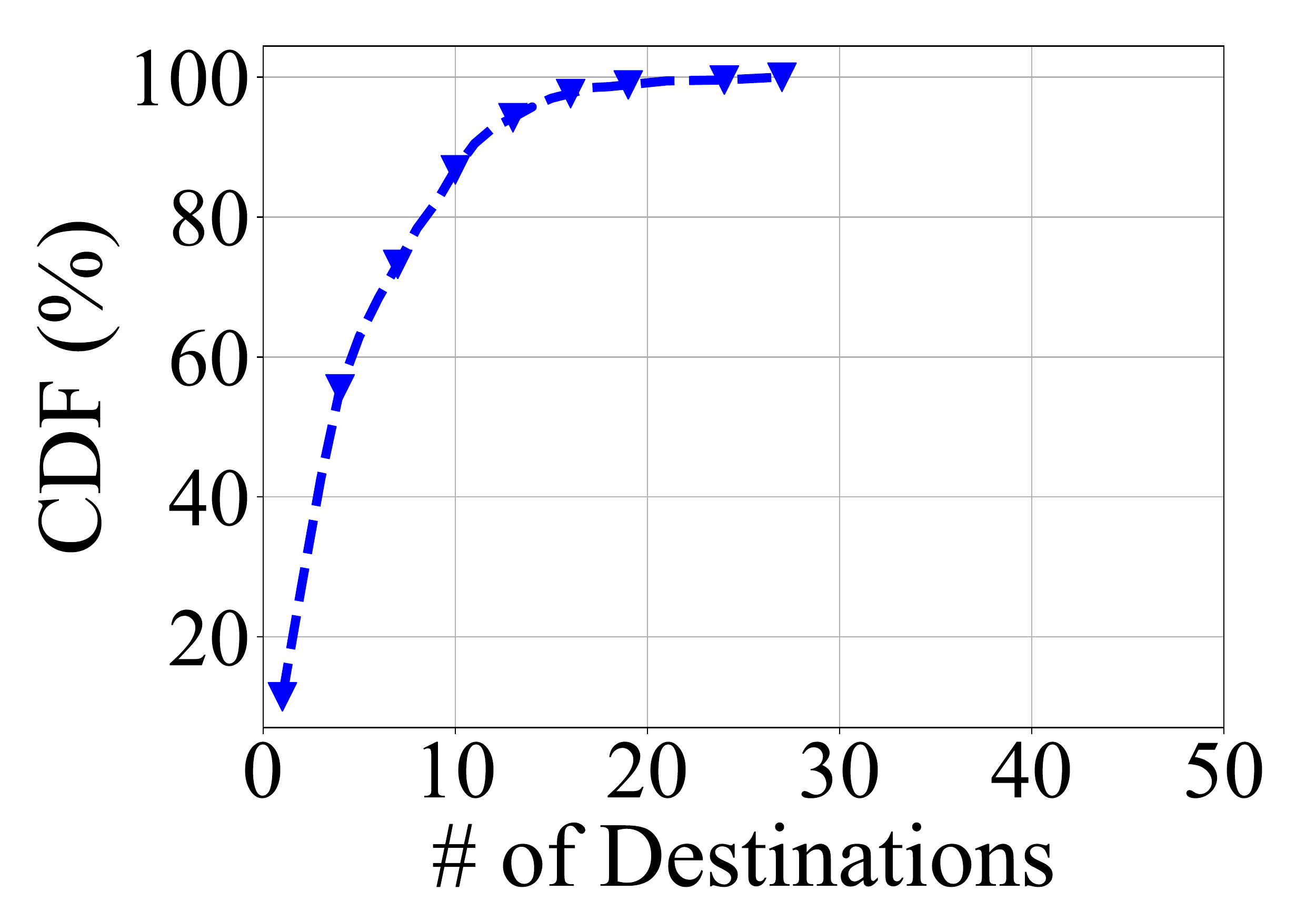}
        % \vspace*{-20pt}
        \caption{Crowd dest.}
        \label{fig:crowd_dest}
    \end{minipage}
    % \vspace*{-5pt}
    \end{figure}

    \textbf{Individual Features:} Since individual destinations essentially are based on personal habits, we utilize a set of individual features.
    \begin{itemize}
        \item \textbf{Historical Destinations:} 
        As shown in Fig~\ref{fig:dest_entropy}, 
        the mobility patterns of most individuals in terms of destinations are relatively stable. Therefore, historical destinations may largely represent their future destinations.
        \item \textbf{Time Factor:}
        Considering the commute pattern in Fig~\ref{fig:dest_entropy} when the entropy is equal to 1, 
        by introducing the entering time factor, the uncertainty of destinations is reduced. 
        We use half-hour time windows to split one day into 48 time slots.
        \item \textbf{Vehicle Type:} 
        It has three values: cars, buses, and trucks. 
        Intuitively, the trucks most probably go to areas with high cargo demand (e.g., industry parks) and buses often go to areas with a dense population (e.g., commercial districts or transportation hubs). Fig~\ref{fig:v_type_dest} shows the proportion of different vehicle types in different types of areas.
        We select three exemplary areas and calculate the proportional of different types of vehicle whose destinations are in the area.
        We found only a few trucks go to the commercial areas; 
        cars and buses contribute major volume in the commercial and transportation hub areas, respectively.
    \end{itemize}
    
    \textbf{Crowd Features:} The individual vehicle's historical data can be very sparse (as we suggested in Section~\ref{sec:moti}). we try to use the crowd destinations to provide complementary information. 
    % Fig~\ref{fig:crowd_dest} shows the possible destinations from the same origins by different percentages of the total transactions (i.e., different crowd sizes). 
    % For example, based on the $30\%$ of transaction data from the same origin (i.e., the 30\% curve), 
    % almost $90\%$ of vehicles go to at most 4 destinations. 
    Fig~\ref{fig:crowd_dest} shows the possible destinations from the same origins by half of all the vehicles.
    We found almost $50\%$ of vehicles go to at most 10 destinations. 
    It indicates lots of vehicles from the same origins share the similar destinations, which can be used to infer the destination of a vehicle without any historical destination data.
    
\begin{figure}[htbp] \centering
    % \vspace*{-5pt}
    \begin{minipage}{0.45\textwidth} \centering
        \includegraphics[width=\textwidth, keepaspectratio=true]{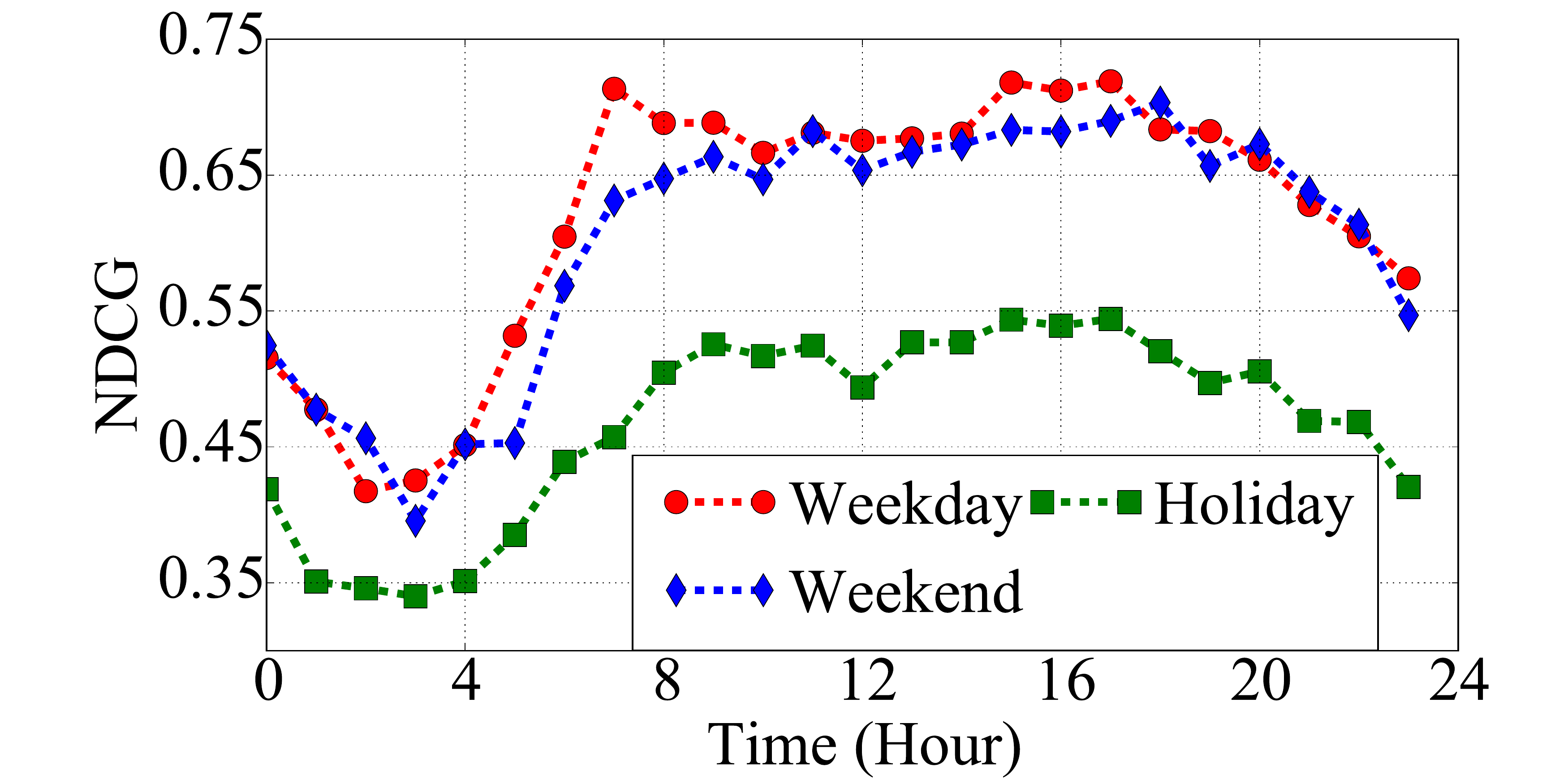}
        % \vspace*{-20pt}
    \end{minipage}
    \caption{Context impacts}
    \label{fig:NDCG}
    % \vspace*{-5pt}
    \end{figure}

    \textbf{Context Features:} We further consider other context features, i.e., the day of the week, weekday/weekend, holidays, that may have impacts on the destination choices. 
    We choose the 10 most popular destinations for each origin and compare the rank of these destinations in a regular day with that in other days with different contexts using the measurement of Normalized Discounted Cumulative Gain $NDCG$~\cite{jarvelin2002cumulated}. 
    The lower the $NDCG$, the lower similarity the destination choices. 
    Fig~\ref{fig:NDCG} shows that the measurement between weekdays, weekend, and holiday. 
    The holiday has very different destination choices compared to other days. 
    In the early morning and the late afternoon of weekends, the $NDCG$ is also lower than that of weekdays. It suggests these factors have impacts on people's choice of the destinations.
    
\subsection{Historical Route and Speed Learning}
\label{sec:historical_data}
As we discuss in Section~\ref{sec:moti}, the reason that previous works are not feasible in our setting is that the historical routes and speeds of individual vehicles cannot be observed by the ETC system. In order to learn the mobility of individual vehicles, we propose a joint learning approach to obtain the historical routes and speeds of vehicles simultaneously, which are utilized as training data to model the route choices and real-time speeds in Section~\ref{sec:route_prediction} and Section~\ref{sec:speed_prediction}.

Several studies~\cite{gao2014elastic}~\cite{yu2016senspeed} have been done to investigate the relationship between the travel routes and real-time speeds, which found the route of vehicles can be inferred with only speeds information. This finding indicates the strong correlation between the routes and speeds, which inspires our idea to learn the routes and speeds simultaneously.

To achieve this, we first present a few preliminaries. 
\begin{itemize}
    \item Time: we divide a day of 24 hours into $K$ time slots($t$) (i.e., each time slot is equal to 10 minutes).
    \item Location: we split the highway road networks into $M$ equal length road segments($s$) (i.e., 1 km).
    \item Speed: instead of treating the speed as a continuous variable, we discretize it into $H$ discrete integer speed($v$) by the smallest unit of 1 km/h (e.g., if the speed limit is 120 km/h, then we can have 121 different speed values ranging from 0 to 120km/h).
\end{itemize}

\begin{figure}[htbp] \centering
    \vspace*{5pt}
    \begin{minipage}{0.35\textwidth} \centering
        \includegraphics[width=\textwidth, keepaspectratio=true]{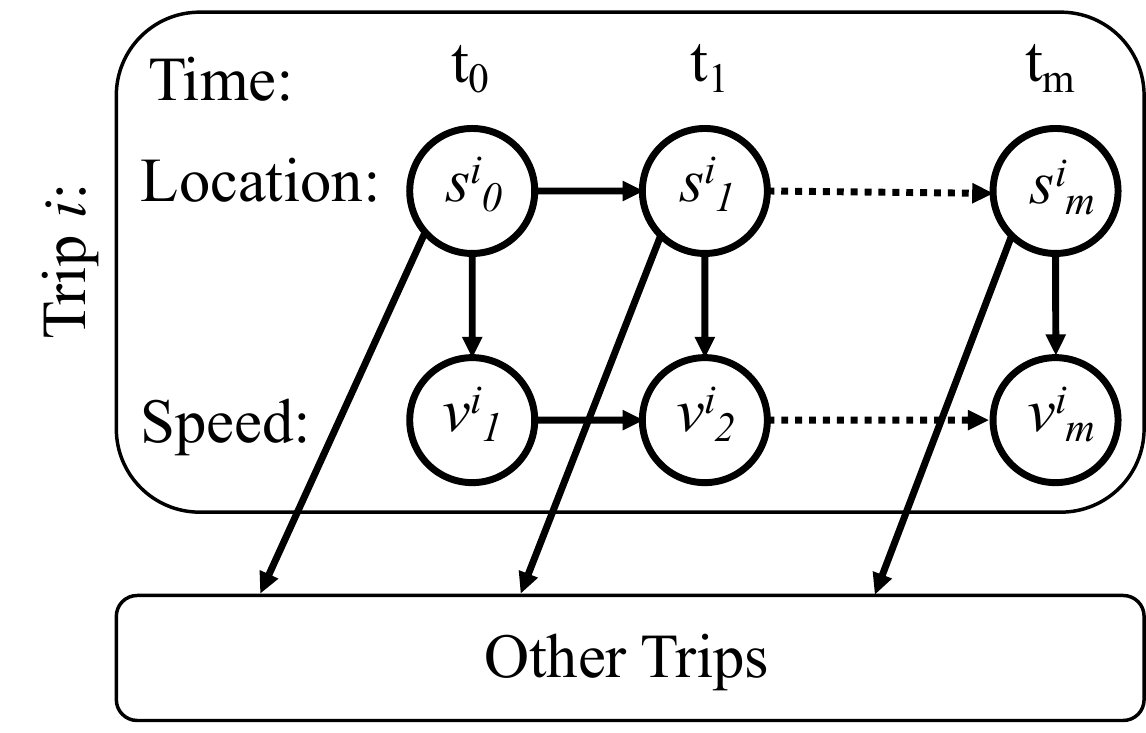}
        % \vspace*{-15pt}
    \end{minipage}
    \caption{Route and speed Correlation}
    \label{fig:route_speed}
    % \vspace*{-5pt}
\end{figure}

\noindent In this way, the states of vehicles in each trip on highways can be presented as a sequence of states <$t$, $s$, $v$> between the origin and the destination.
As an example of the trip $i$ in Fig~\ref{fig:route_speed}, the vehicle enters the highway from the road segment $s_0$ at the time $t_0$ and exits the highway from the road segment $s_m$ at the time $t_m$. 
It is worth mentioning that, in other trips, vehicles can be at the same location as the same time as the trip $i$.
Then our objective is to infer the most likely state sequence of each trip.
The solution is motivated by the key observation that at the same time multiple vehicles are traveling on the same road segments and their real-time speeds can be considered as samples of the speed distribution. The following insights reveal the characteristics of the distribution.
\begin{itemize}
    \item \textbf{Speeds distribution on the road segment}: By analyzing the sample GPS trajectories, we observe that speeds of vehicles on the same road segment follow a normal distribution, which is also validated in other contexts~\cite{hellinga2008decomposing}.
    \item \textbf{Speed STD distribution}: Moreover, as shown in Fig~\ref{fig:speed_std}, we also observe strong normality of the speed.
\end{itemize}
Since both insights show the normality, to quantify them, we utilize Kolmogorov-Smirnov test to test the normality.
Specifically, the states of different trips within the same time and location are grouped as samples to test the normality of speed on the road segments.
For the speed STD distribution insight, it is measured as suggested in Section~\ref{sec:moti}.
Then all the STDs are considered as samples to test the normality.

Given the normality test of both the speed distribution in each road segment and speed STD distribution of all the vehicles, our problem can be transformed into an optimization problem to find the best state sequence combination for the maximization of the number of the acceptance of normality tests.
Suppose we have $N$ trips with $J$ vehicles, we formulate the problem as following:
\begin{equation*}
\begin{aligned}
& \underset{sc}{\text{maximize}}
& & \sum_i^{N}\mathbbm{1_A}{(Rnorm(sc))} + \sum_j^{J}\mathbbm{1_A}{(Snorm(sc))}
\end{aligned}
\end{equation*}
where $sc$ is the combination of the state sequences of different trips, $Rnorm$ is a test function to check the normality of the speed distributions, $Snorm$ is a test function to check the normality of the speed STD distribution. $\mathbbm{1_A}$ is an indicator function of the test acceptance.

A straightforward approach to solve the optimization problem is to search all the possible state sequence combinations. For each trip, the possible state sequence is $K\times M \times H$. Then the total search space is $O(N^{K\times M \times H})$, which is time consuming to search. To reduce the search space, we introduce several simple but effective heuristics to guide the search.
\begin{itemize}
    \item State sequences constrained by routes: Shown in Fig~\ref{fig:num_routes}, there is a limited number of routes between origins and destinations, which naturally reduces the search space of possible location sequences.
    
    \item Spatial smoothness: Constrained by the structure of the road network and the speed limit, the next location of the vehicle can be the reachable road segments under the speed limit. (e.g., suppose the speed limit is 120km/h, the next location in 5 minutes can only be the road segments within a range of 5 minutes$\times$120 km/h = 10 km.)
\end{itemize}
\noindent Given these heuristics, we perform a standard search algorithm (e.g., DFS) to find the best combination of the state sequence. Then the historical routes can be obtained by concatenating the locations in each trip and speeds can be directly obtained from the state sequence.

\subsection{Route Predictor}
\label{sec:route_prediction}

Similar to the destination prediction, we study the features from three perspectives: individual features, crowd features and context features.

\textbf{Individual Features:} We utilize the following features for the route prediction at the individual level. 
    \begin{itemize}
        \item \textbf{Historical Routes:} 
        Based on a previous study, people are more reluctant to change their regular routes if they have more experience with these routes~\cite{ben2010road}, which indicates historical routes are most likely to be their future routes given the same origin and destination.
        \item \textbf{Driving Experience:} Empirically, experienced people are good at finding the best routes~\cite{ben2010road}.
        We quantify the experience by two factors: 
        (i) the frequency of driving on highways, which can be obtained from historical ETC transactions; (ii) the saved travel time compared with the average travel time, which can also be computed from historical ETC data.
        \item \textbf{Time Factor:} Empirically, people generally have their own estimations about the route traffic at a different time, e.g., taking a detour during the rush hour to avoid the traffic. It affects their future route choices.
    \end{itemize}
    
    \textbf{Crowd Features:}
    For those people who have no or only limited historical data, we incorporate the route choices of crowds to infer their route choice. Specifically, we use the probability of historical crowds' routes between particular origin/destination at the certain time.

    \textbf{Context Features:} 
    People's route choices are affected by the real-time context~\cite{ben2013impact}, i.e., the day of the week and real-time traffic speed, which can be estimated with ETC transactions in the recent past.

\subsection{Speed Predictor}
\label{sec:speed_prediction}
In this subsection, we introduce different features that are correlated to the real-time speed. 
% MODIFY
The key idea is to learn the relation between individual driving speed and other features (e.g., crowd speed) in order to predict the real-time speed given all these features.

\subsubsection{\textbf{Features:}} 
\label{sec:indi_speed}
We introduce our features on the individual, crowd, and context level.

\begin{figure}[htbp] \centering
        % \vspace*{-5pt}
        \begin{minipage}{0.5\linewidth} \centering
    %         \hspace*{-30pt}
            \includegraphics[width=\linewidth, keepaspectratio=true]{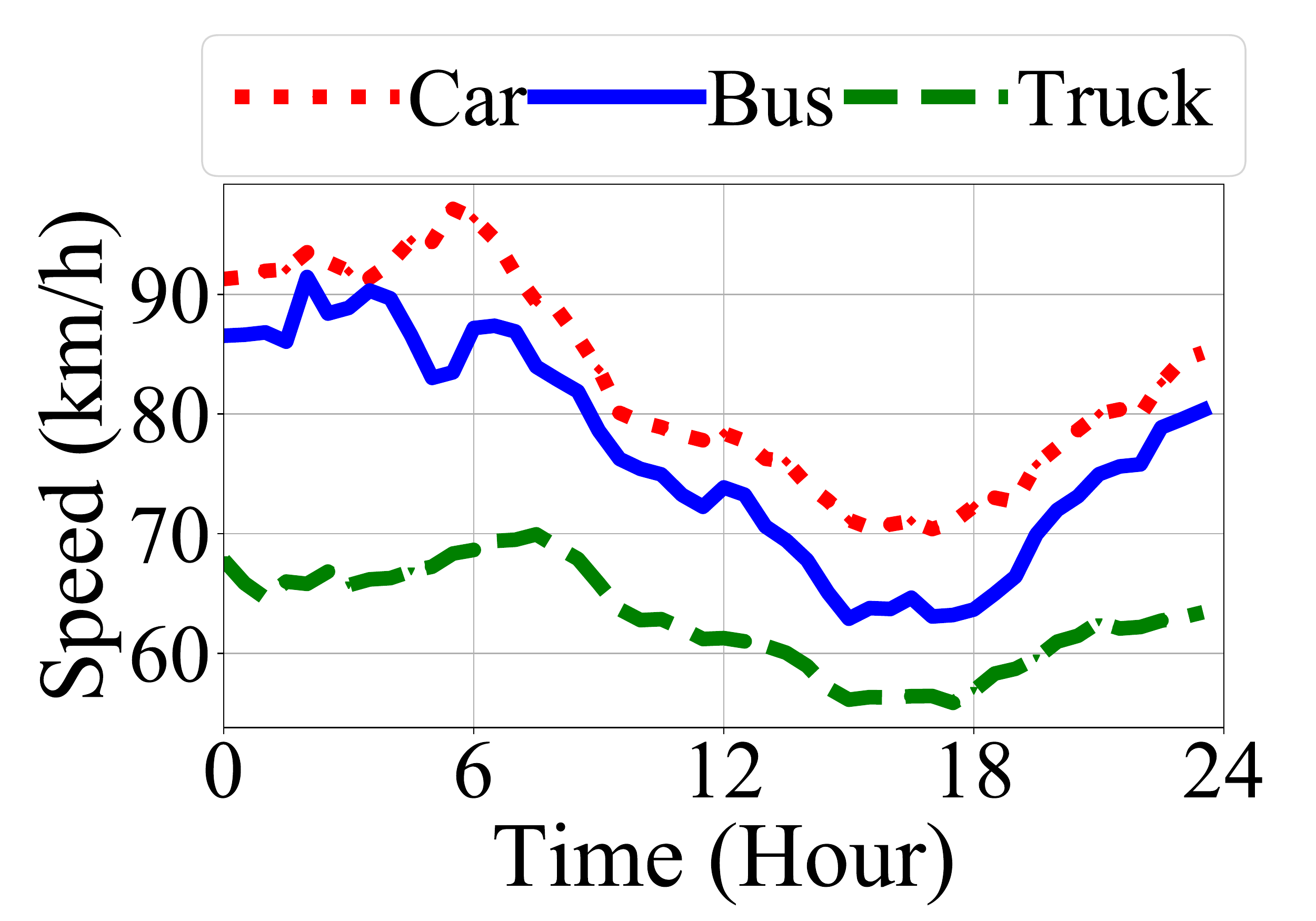}
            % \vspace*{-20pt}
            \caption{Speed variance}
            \label{fig:v_type_speed}
            % \hspace*{5pt}
        \end{minipage}
        \hspace*{-5pt}
        \begin{minipage}{0.5\linewidth} \centering
    %         \hspace*{-30pt}
            \includegraphics[width=\linewidth, keepaspectratio=true]{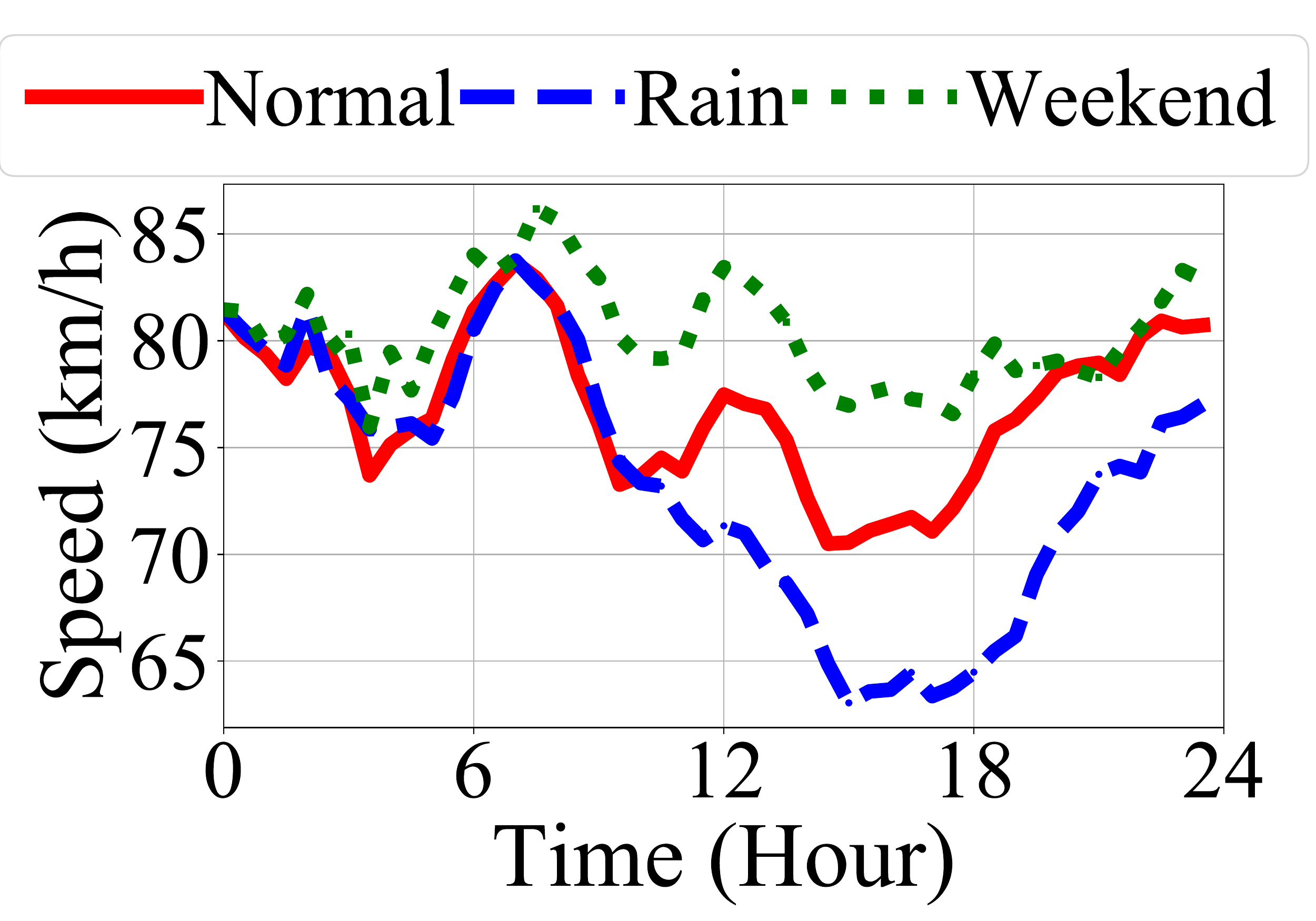}
            % \vspace*{-20pt}
            \caption{Context impacts}
            \label{fig:context_speed}
        \end{minipage}
        % \vspace*{-5pt}
    \end{figure}

     \textbf{Individual Features:} Since the driving speed is essentially based on people's behaviors, we define a set of individual vehicle's features.
    \begin{itemize}
        \item \textbf{Historical Driving Speed:} As shown in Fig~\ref{fig:speed_std}, the driving speed is relatively stable for a particular person. 
        We use their average speeds of historical trips to reflect their general driving speed.
        
        \item \textbf{Vehicle Type:} This feature reflects the vehicle's type (i.e., cars, buses, trucks). Intuitively, the driving speed of cars should be higher than trucks and buses. Fig~\ref{fig:v_type_speed} also validates this intuition.
        
        \item \textbf{Time Factor:} Fig~\ref{fig:v_type_speed} shows that the driving speed varies at the different time of a day, which is mainly due to the different traffic conditions.
    \end{itemize}

\begin{figure}[htbp] \centering
    % \vspace*{-5pt}
    \begin{minipage}{0.45\textwidth} \centering
        \includegraphics[width=\textwidth, keepaspectratio=true]{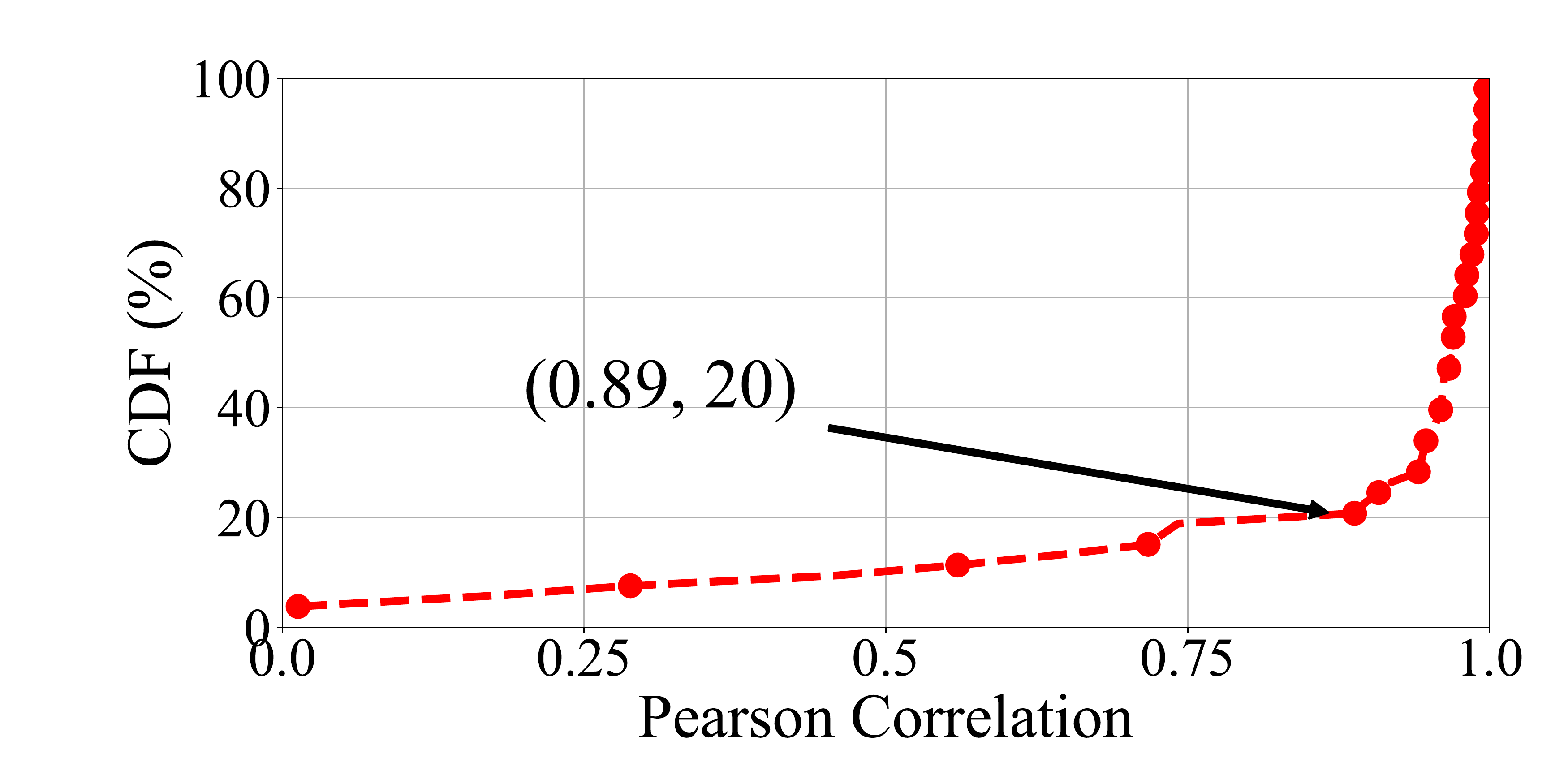}
        % \vspace*{-20pt}
    \end{minipage}
    \caption{Speed Correlation}
    \label{fig:speed_cor}
    % \vspace*{-5pt}
    \end{figure}

    \textbf{Crowd Features:} People may behave differently under different traffic conditions. Instead of studying the detailed behavior patterns of individuals, which may have many factors to discuss, we directly investigate the correlation between the individual speed and crowd speed. Fig~\ref{fig:speed_cor} shows the Pearson correlation of the individual speed and crowd traffic speed. More than 80$\%$ of vehicles have at least 0.89 correlation coefficient with the crowd traffic speed. Motivated by the strong correlation, the crowd traffic speed is an important feature to estimate the individual driving speeds on specific edges.
    Therefore, we extract the features of the vehicle speed samples, which are incorporated to estimate the crowd traffic speed. Instead of using the average crowd traffic speed (which may cause an estimation bias), we consider the statistic values of the crowd traffic speed distribution, including minimum, lower fourth, median, upper fourth and maximum of the samples.
    We reply on the crowd features to learn how the driver would react under different situations, in order to predict the real-time speed in the future.
        
    \textbf{Context Features:} Besides the vehicle-related features, we also consider other factors that may have impacts on the driving speed, including weather and weekday/weekend. As shown in Fig~\ref{fig:context_speed}, the speed is decreased by $10\%$ at most in the rainy day and increased by $5\%$ on weekend. This is reasonable because people tend to drive slower when raining and fewer people use highways to work on the weekend, which makes the highways less congested.

\subsection{Learning with Mondrian Forest}
Mondrian forests~\cite{lakshminarayanan2014mondrian} is an online random forest model using Mondrian processes to construct ensembles of decision trees.
Compared to the offline or online random forest~\cite{lakshminarayanan2014mondrian}, it provides the ability to process online data and online updates faster and more accurately.
Compared with other algorithms, the Mondrian forests model has the following advantages:
\begin{itemize}
    \item It is more robust to heterogeneous features. In our data input, we have both numerical variables (i.e., speed) and categorical values (i.e., vehicle type, weather, weekday/weekend). These variables can be input into the model directly without conversion or normalization.
    \item It provides self-check on the importance of the features during the training stage. For example, such as the weather condition and holidays, these variables would only have high importance under certain conditions with a low frequency.
    %\item It is efficient for large data input. Considering the huge volume in ETC systems, it ensures the real-time effects of the estimation.
    \item Compared to other neural based model (e.g., deep neural network), the results are more explainable because of the internally used decision tree~\cite{friedman2001elements}.
\end{itemize}

For different tasks (i.e., destination prediction, route inference, speed estimation), we fit all the extracted features into Mondrian forests and learn three predictors to work collaboratively on the real-time location prediction, which is illustrated in Section~\ref{sec:together}.
% MODIFY
Even we choose Mondrian forests, our system is flexible to many machine learning methods. The more important aspect is the analysis process and find the effective features. 

\subsection{Put them all together}
\label{sec:together}
In the previous sections, we have conducted an analysis of the three key tasks: destination prediction, route inference and speed estimation.
Based on multiple extracted features, we learn three predictors \textit{d-predictor, r-predictor, s-predictor} for each of the tasks, perceptively. The procedure of real-time location prediction is described in Algorithm~\ref{alg:algo}.

\begin{algorithm}
\caption{Real-time Location Prediction}\label{alg:algo}
\SetKwInOut{Input}{Input}
\SetKwInOut{Output}{Output}
\Input{\textit{d-predictor}: the destination predictor,\\
        \textit{r-predictor}: the route predictor,\\
        \textit{s-predictor}: the speed predictor,\\
        $entrance$: the entering toll station,\\
        $interval$: the updating time interval\\
        $t_0$: the entering time.}
\Output{real-time locations}
$destination\leftarrow$ \textit{d-predictor} given $entrance$\\
$route\leftarrow$ \textit{r-predictor} given $destination$\\
$distance$ = 0\\
\While{$distance$ < $route.length$}{
    $speed\leftarrow$ \textit{s-predictor} at $t_i$\\
    $distance$ += $speed\times interval$\\
    $location\leftarrow$ match $distance$ to $route$\\
}
\end{algorithm}
% \vspace{-10pt}

\section{Evaluation}
\label{sec:eval}

In this section, we introduce our data-driven evaluation in terms of methodology and results. 

\subsection{Evaluation Methodology}
\begin{figure}[htbp] \centering
    % \vspace*{-10pt}
    \begin{minipage}{0.48\textwidth} \centering
        \includegraphics[width=\textwidth, keepaspectratio=true]{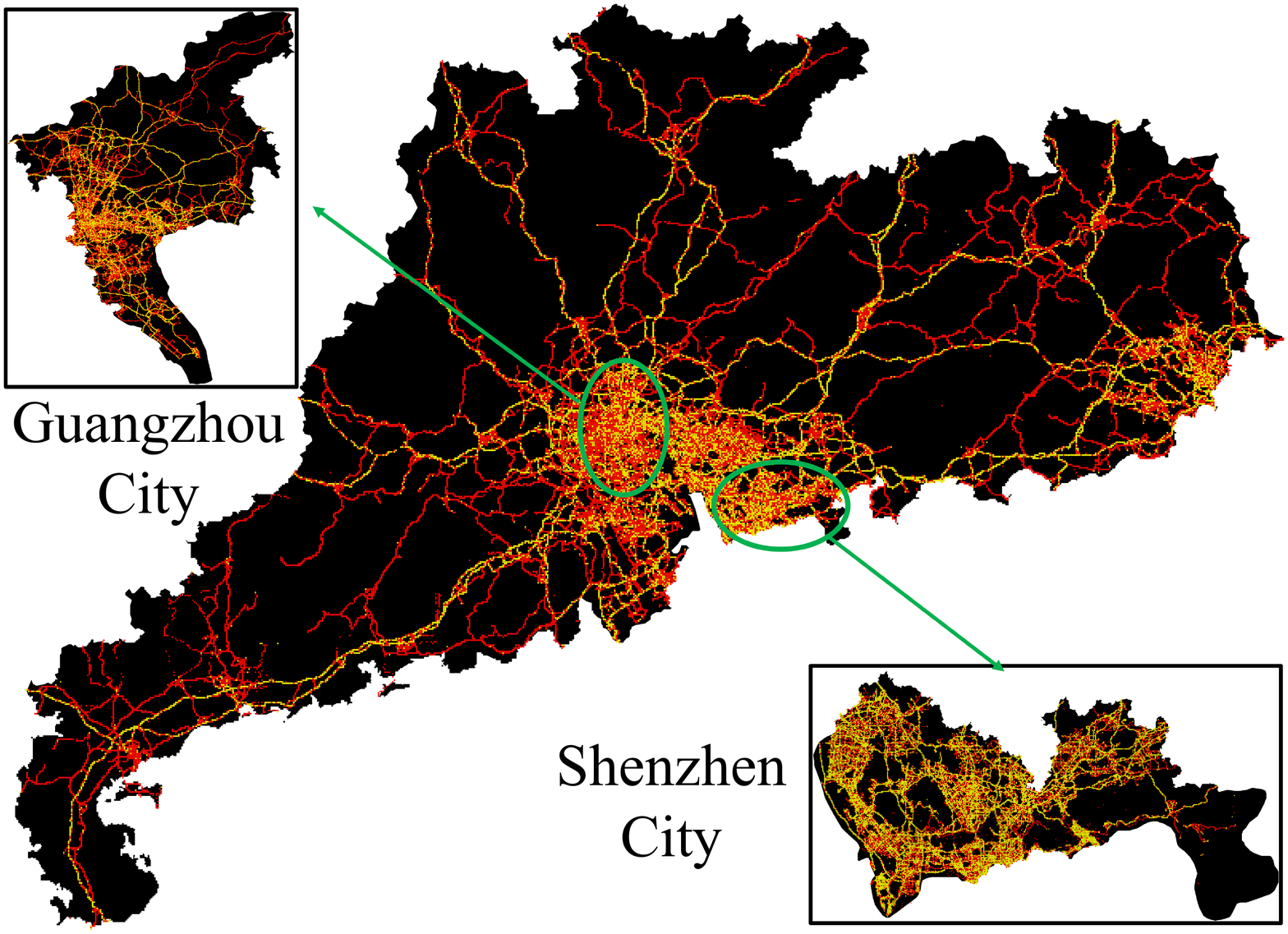}
        % \vspace*{-20pt}
    \end{minipage}
    \caption{Ground truth visualization}
    \label{fig:gps}
    % \vspace*{5pt}
\end{figure}
\noindent\textbf{Ground Truth:} To obtain the ground truth of real-time vehicle locations, we introduce another real word dataset with detailed GPS trajectories in Guangdong, which provide the real-time locations of 114 thousand vehicles including 75\% cars, 13\% buses and 12\% trucks.
These vehicles upload their real-time locations in every 10 to 30 seconds. The detailed data format is presented in Table~\ref{tb:truth}. Fig~\ref{fig:gps} shows the trajectories visualization on the main roads in Guangdong.
It shows our dataset covers most of the main roads, which can be utilized to evaluate the state-level mobility.
Two highlighted areas are the two largest cities, i.e., Guangzhou and Shenzhen, which are densest areas in terms of vehicles.
\noindent For each vehicle, we first apply a map matching algorithm~\cite{newson2009hidden} to map trajectories onto the road network.
Then only the trajectories on highways are remained to obtain entering toll stations, exit toll stations, routes and real-time locations, which cover 20\% of the vehicles in our dataset. Since the training and testing are conducted on different datasets, we do not need to split the datasets for cross-validation.

\begin{table}[htbp]
% \vspace*{-5pt}
\centering
\caption{Ground Truth Format}
\label{tb:truth}
\begin{tabular}{@{}llll@{}}
\toprule
Field       & Value                 & Field      & Value        \\ \midrule
Id          & P0SF51B4GU            & Type       & Car/Bus/Truck          \\
Longitude   & 113.402904            & Latitude   & 23.167894    \\
Time        & 2016-06-01 00:00:34   &            &              \\ \midrule
% \multicolumn{4}{l}{Date: 2016-06-01 to 2016-06-30}              \\
\multicolumn{2}{l}{75\% cars, 13\% buses, 12\% trucks}  & \multicolumn{2}{l}{\#Vehicle: 114k} \\
\bottomrule
\end{tabular}
\end{table}

\noindent\textbf{Evaluation Metrics:}
For each component, we define the evaluation metrics as follows:
\begin{itemize}
    \item Destination and Route prediction:
    \begin{equation}
        accuracy=\frac{\#prediction_{correct}}{\#prediction_{all}}\times 100\%
    \end{equation}
    where $\#prediction_{correct}$ is the number of corrected prediction and $\#prediction_{all}$ is the total number.
    
    \item Speed Prediction: 
    \begin{equation}
        accuracy=1 - \frac{|speed_{predict} - speed_{actual}|}{speed_{actual}}
    \end{equation}
    where $speed_{predict}$ is the predicted speed and $speed_{actual}$ is the ground truth.
    
    \item Real-Time Location Prediction: we quantify the location accuracy by measuring the percentage of predicted locations within the accuracy threshold (i.e., 100 meters) of the ground truth considering the GPS errors every 15 seconds (i.e., the average uploading time interval of data in ground truth)~\cite{grewal2007global}. The accuracy formula is defined as
    \begin{equation}
    \label{eq:correct}
        accuracy=\frac{\#prediction_{correct}}{\#prediction_{all}}\times 100\%
    \end{equation}.
\end{itemize}

\noindent\textbf{Baselines for Intermediate Results:}
For the three individual prediction components, i.e., predictions for destinations, routes, and speeds, since we utilize a unified algorithm for all of them, we evaluate them from the perspective of a learning model by comparing it with the other learning models. The selected learning models are presented as follows, and each of them is representative of a group of methods with the similar bases:
\begin{itemize}
    \item \textbf{Empirical Estimation (Emp)}: The baseline represents the prediction based on the naive empirical knowledge. For the destination and route prediction, we consider the most frequently visited destinations and routes. For speed prediction, we utilize their historical average speed.
    \item \textbf{Bayesian Network (Bayes)}~\cite{friedman2001elements}: Bayesian network is a typical graph-based algorithm, which is representative for the probability based models.
    \item \textbf{Neural Network (Neural)}~\cite{friedman2001elements}: Neural network represents the models that focus on learning the linear or non-linear combination between features and targets.
\end{itemize}

\noindent\textbf{Baselines for End-to-End Results:}
For the overall performance of the real-time locations, we choose the baselines based on two principles: (i) static infrastructure based methods; (ii) mobile sensor based methods.
\begin{itemize}
    
    \item \textbf{STrack}: This baseline represents a wide range of static infrastructure based methods, e.g., cameras~\cite{zhang2017fcn}. Considering traffic cameras are set to detect motoring offenses without open location information, we implement STrack by assuming a given percentage of edges (defined in Section~\ref{sec:infras}) have been installed with cameras that can track vehicles. In the middle of each edge, we assume one traffic camera is installed that can recognize vehicle plates. The real-time locations of vehicles are obtained as being observed by the cameras. For the location estimation of vehicles between cameras, we assume they are uniformly distributed on the roads between cameras. Fig~\ref{fig:edge_length} shows the edge length in the highway road network. Different percentages are also evaluated to show the performance.
    
    \begin{figure}[!htbp] \centering
    % \vspace*{-5pt}
    \begin{minipage}{0.46\textwidth} \centering
        \includegraphics[width=\textwidth, keepaspectratio=true]{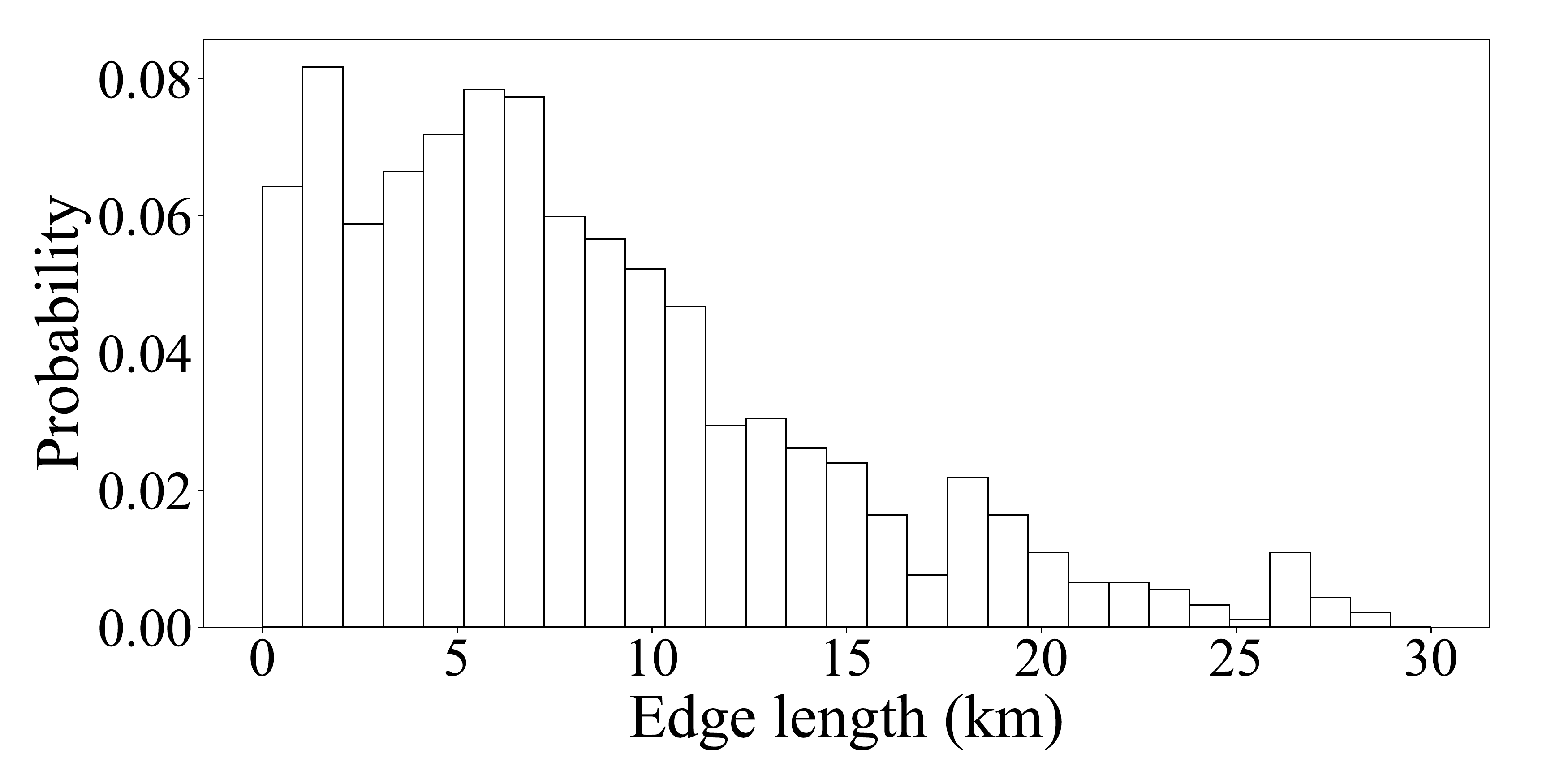}
        % \vspace*{-20pt}
    \end{minipage}
    \caption{Edge length}
    \label{fig:edge_length}
    % \vspace*{-5pt}
    \end{figure}

    \item \textbf{CTrack}~\cite{thiagarajan2011accurate}: This baseline aims to track individual vehicles based on cellular networks by periodical communications between onboard cellphones and cell towers. Based on the locations of communicated cell towers, it infers the locations of the cellphones (thus vehicles). The cell tower locations we use are located in Shenzhen City (shown in Fig~\ref{fig:cell_location}), where the ETC system is also widely spread with 79 toll stations. We implement CTrack by assuming each vehicle has an onboard cellphone to interact with cell towers and follow the trajectory mapping algorithm in ~\cite{thiagarajan2011accurate}.
    
    \begin{figure}[htbp] \centering
    \vspace*{5pt}
    \begin{minipage}{0.48\textwidth} \centering
        \includegraphics[width=\textwidth, keepaspectratio=true]{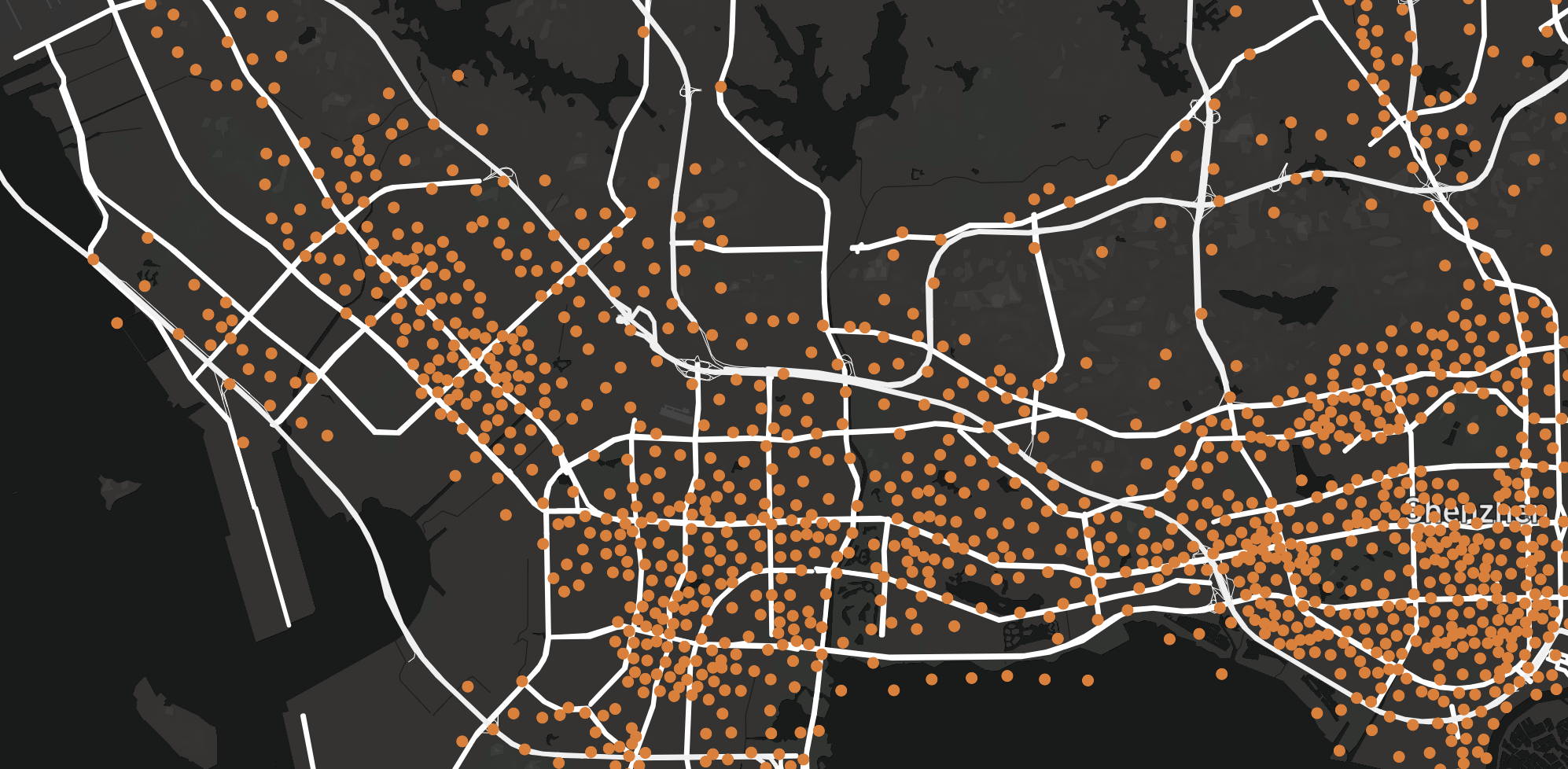}
        % \vspace*{-20pt}
    \end{minipage}
    \caption{Cell Tower Locations in Shenzhen}
    \label{fig:cell_location}
    % \vspace*{-5pt}
    \end{figure}
    
\end{itemize}

\noindent\textbf{Impacts of Factors:} We evaluate several factors to show the impacts on the performance of VeMo, 
\begin{itemize}
    \item \textbf{Weather}: Weather condition is a factor that affects the driving behavior such as driving speed. We evaluate the accuracy in both regular day and extreme weather day (e.g., heavy rain).
    \item \textbf{Accuracy threshold}: Given different accuracy threshold to declare the accuracy, the performance may be varied. We choose several threshold values to show the accuracy changes.
    \item \textbf{Time factors:} The performance may vary at different time. We evaluate it in weekday, weekend and holiday.
    \item \textbf{Spatial factors:} Shown in the previous sections, different areas have different densities of toll stations and different volume of traffic. We evaluate VeMo at different areas in Guangdong, i.e., both the downtown areas and suburb areas.
    \item \textbf{Vehicle Types:} Different types of vehicles may have different challenges of prediction. We evaluate this factor by applying VeMo on different types of vehicles, i.e., car, bus, truck.
\end{itemize}

In the next subsection, we first compare the three individual components with the baselines (i.e., Naive Empirical, Bayesian, Neural). Then we comparing the overall performance of the real-time location prediction with \textit{STrack} and \textit{CTrack} followed by the impacts of the factors.

\noindent\subsection{Evaluation Results}

\subsubsection{Efficiency}

We implement VeMo on a server with Intel Xeon E5-1660 3.00GHz CPU and 32GB RAM in 16 threads. After loading all the data, the training process takes 450 seconds. The speed prediction is 500 times per thread every second on average, which can satisfy the real-time need of 4 million daily transactions.

\subsubsection{Real-Time Edge-Cloud Design}
Since most of the applications built on our system require real-time response, it is necessary to have real-time cloud components. 
Even it is feasible to conduct prediction in a powerful server, however, it is challenging to update the model in the cloud in real-time. Our solution is to combine both the cloud (i.e., center servers) and the edges (i.e., computer systems in the toll stations). \\
\textbf{Cloud:} 
All the data is stored in the cloud system for security issues.
As the new data collected in the edges, the data is transmitted to the cloud through Ethernet.
All the trained models are also stored in the cloud to distribute to the edges.\\
\textbf{Edge:}
Given the truth that a vehicle only appears in a few toll stations, we could pre-distribute the trained individual models to top frequent edges according to historical records. 
Considering the online updating feature of our model, we update the model directly in the edge devices.
Then the model itself is transmitted back the cloud to distribute to other station. 
Generally, a vehicle leaving the toll station would not get back to the highways immediately. 
There there is enough time to transmit the model to the cloud.

\subsubsection{Comparison to baselines} We evaluate both individual predictors and overall location predictor. For each of the individual component, we evaluate it by comparing it to the three baselines, respectively. Then three predictors work collaboratively to predict the locations of vehicles.

% \noindent\textbf{Individual predictors' performances:}

\begin{figure}[htbp] \centering
    \vspace*{5pt}
    \begin{minipage}{0.5\linewidth} \centering
        \includegraphics[width=\linewidth, keepaspectratio=true]{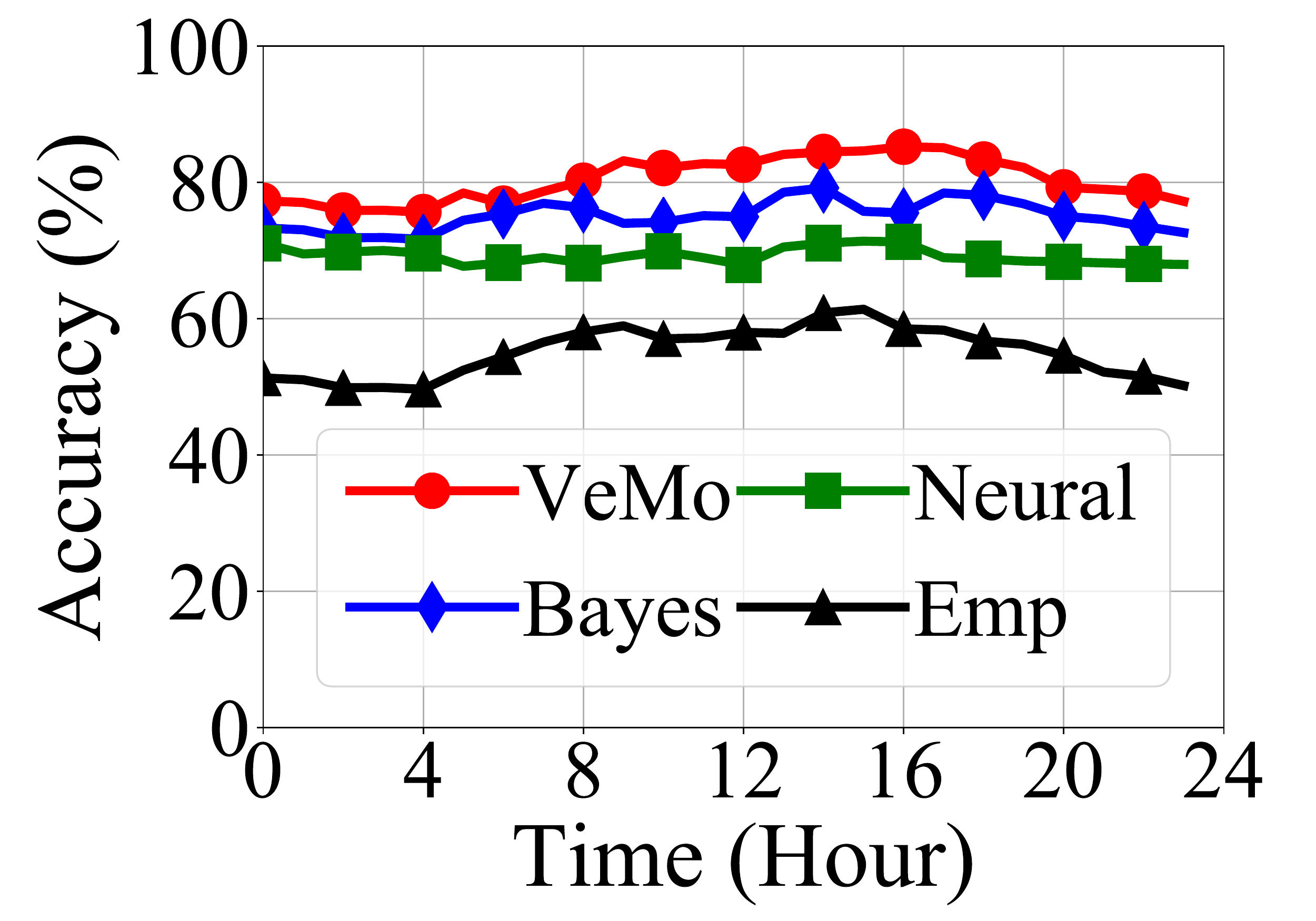}
        % \vspace*{-20pt}
        \caption{Destination pred.}
        \label{fig:evl_dest}
        % \hspace*{5pt}
    \end{minipage}
    \hspace*{-5pt}
    \begin{minipage}{0.5\linewidth} \centering
        \includegraphics[width=\linewidth, keepaspectratio=true]{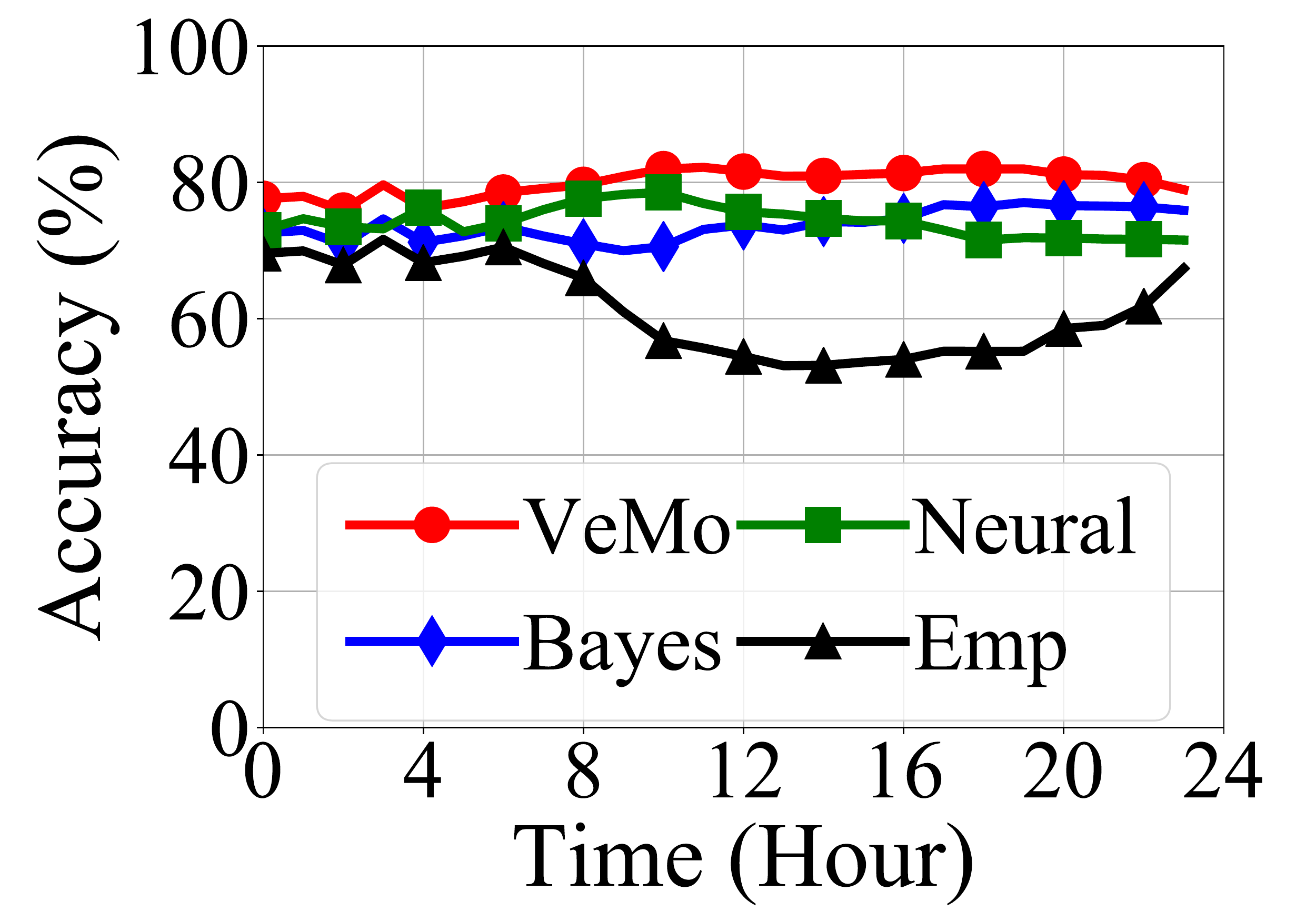}
        % \vspace*{-20pt}
        \caption{Route pred.}
        \label{fig:evl_route}
    \end{minipage}
    % \vspace*{-5pt}
\end{figure}

\noindent\textbf{(i) Destination prediction:} Fig~\ref{fig:evl_dest} plots the result of the destination prediction. 
It shows VeMo has better performance than other three learning models with an average performance gain of $11\%$. 
The Bayesian network baseline performs better than the neural network, which means the probability relationship is better to model the destination prediction problems.
Moreover, the naive empirical baseline achieves $60\%$ accuracy during the day time, which suggests the destination choices are relatively stable on highways.

\noindent\textbf{(ii) Route prediction:} Fig~\ref{fig:evl_route} presents the result of the route prediction.
Compared to the other three baselines, VeMo achieves an average performance gain of $6\%$.
It suggests the performance does not vary much in terms of different learning models.
The naive empirical baseline has similar performance in the early morning but poor performance during the daytime, which means the route choices are flexible when there is heavy traffic.

% Compared to destination prediction, all the methods have relatively worse performance.
% One important reason that makes the route prediction problem harder is that the training data is labeled by estimated routes rather than ground truth, which makes the learned predictor hard to cover all the choices.

\begin{figure}[htbp] \centering
    \begin{minipage}{0.5\linewidth} \centering
%         \hspace*{-30pt}
        \includegraphics[width=\linewidth, keepaspectratio=true]{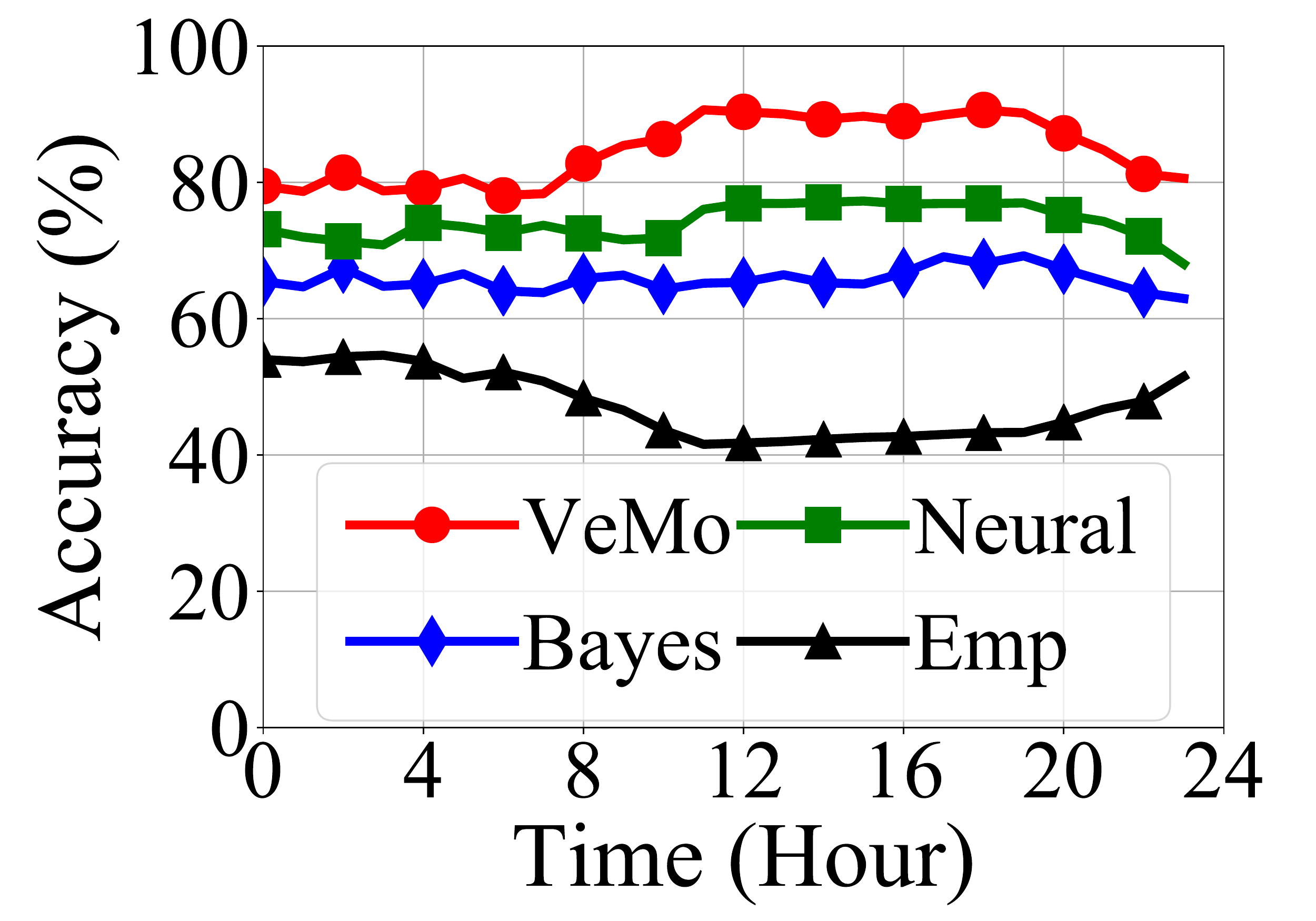}
        % \vspace*{-20pt}
        \caption{Speed pred.}
        \label{fig:evl_speed}
        % \hspace*{5pt}
    \end{minipage}
    \hspace*{-5pt}
    \begin{minipage}{0.5\linewidth} \centering
%         \hspace*{-30pt}
        \includegraphics[width=\linewidth, keepaspectratio=true]{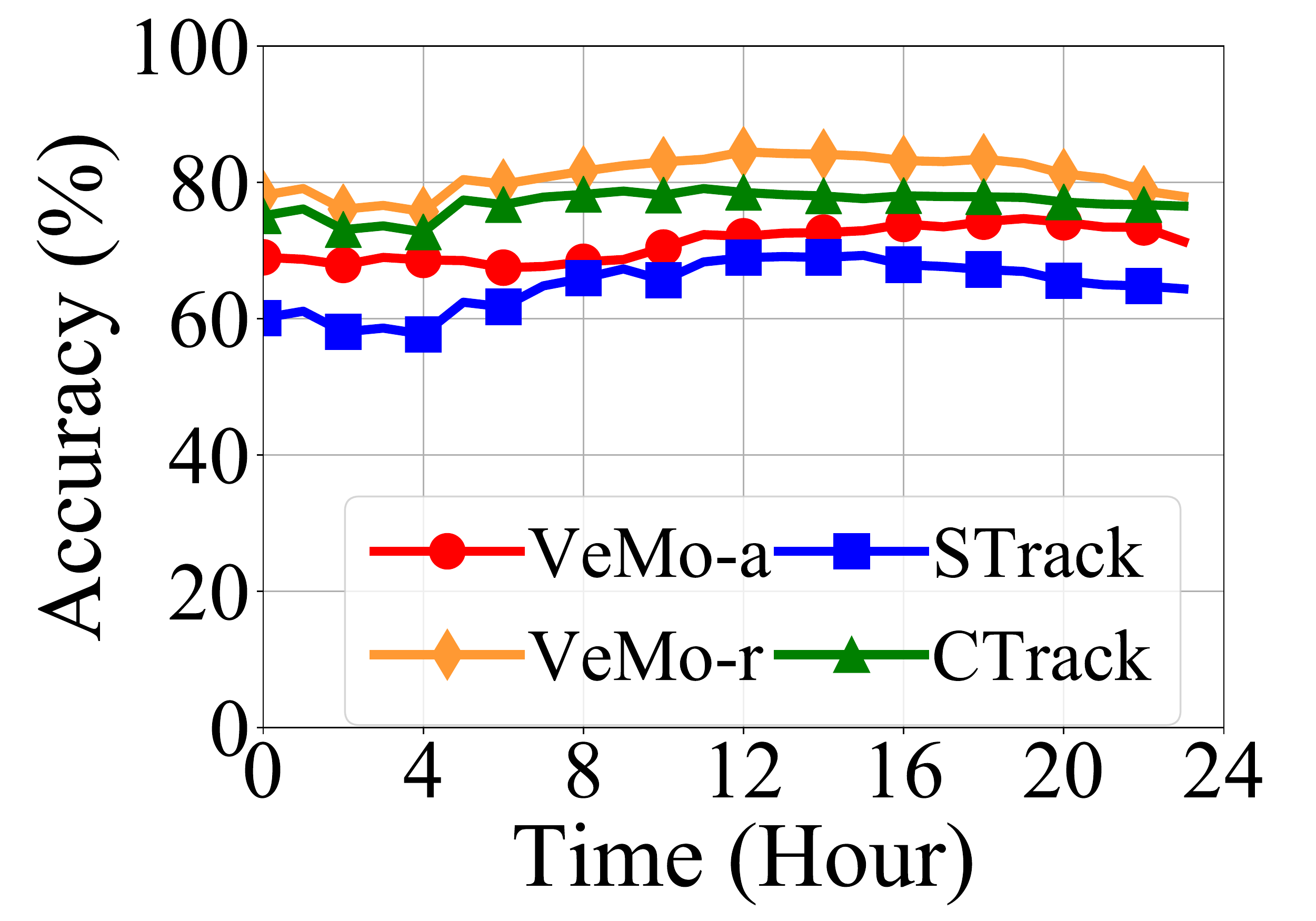}
        % \vspace*{-20pt}
        \caption{Location pred.}
        \label{fig:evl_location}
    \end{minipage}
    % \vspace*{-5pt}
\end{figure}

\noindent\textbf{(iii) Speed prediction:}
Fig~\ref{fig:evl_speed} shows the result of average speed prediction. VeMo has an average performance gain of $17\%$. During the day time, the accuracy is higher, because the heavy traffic constrains the speed variation. The naive empirical baseline shows poorer performance during the daytime because the empirical knowledge cannot obtain the real-time traffic information. Moreover, the neural network baseline is better than the Bayesian-based baseline, which suggests the advantage of linear combination based method on the speed prediction tasks.

\noindent\textbf{(iv) Location prediction performance:}
After the individual predictors' evaluation, we combine them together to evaluate the real-time locations of vehicles. 
Since the route has dominating impacts on the locations of vehicles, to show more sophisticated evaluations, we test the accuracy of both (i) the vehicles (\textit{VeMo-a}) and (ii) those vehicles with correctly predicted routes (\textit{VeMo-r}).
Then we compare them with \textit{STrack} and \textit{CTrack}.
Fig~\ref{fig:evl_location} plots the evaluation results. 
Considering the vehicles with correctly predicted routes, VeMo (shown as \textit{VeMo-r}) has the average accuracy about $82\%$. The reason that VeMo has similar accuracy as \textit{CTrack} is that the baseline experiment is conducted inner city, which has a dense cell tower distribution. Even including all the vehicles (shown as \textit{VeMo-a}), VeMo achieves average accuracy of $70\%$, which is still at the same level of \textit{STrack}, which means VeMo can be an alternative solution of \textit{STrack} without introducing extra infrastructures.

We also evaluate the impacts of coverage percentage of \textit{STrack}, and show the result in Fig~\ref{fig:evl_strack}. After the coverage percentage increases to $50\%$, \textit{STrack} achieves better performance. Since it is expensive to provide such high infrastructure coverage, VeMo outperforms \textit{STrack} in terms of feasibility.

\subsubsection{Impacts of factors}Five factors are evaluated including accuracy threshold, weather, time factors, spatial factors and vehicle types. The metrics are the same as the equation~\ref{eq:correct}.

\begin{figure}[htbp] \centering
    % \vspace*{-5pt}
    \begin{minipage}{0.5\linewidth} \centering
        \includegraphics[width=\linewidth, keepaspectratio=true]{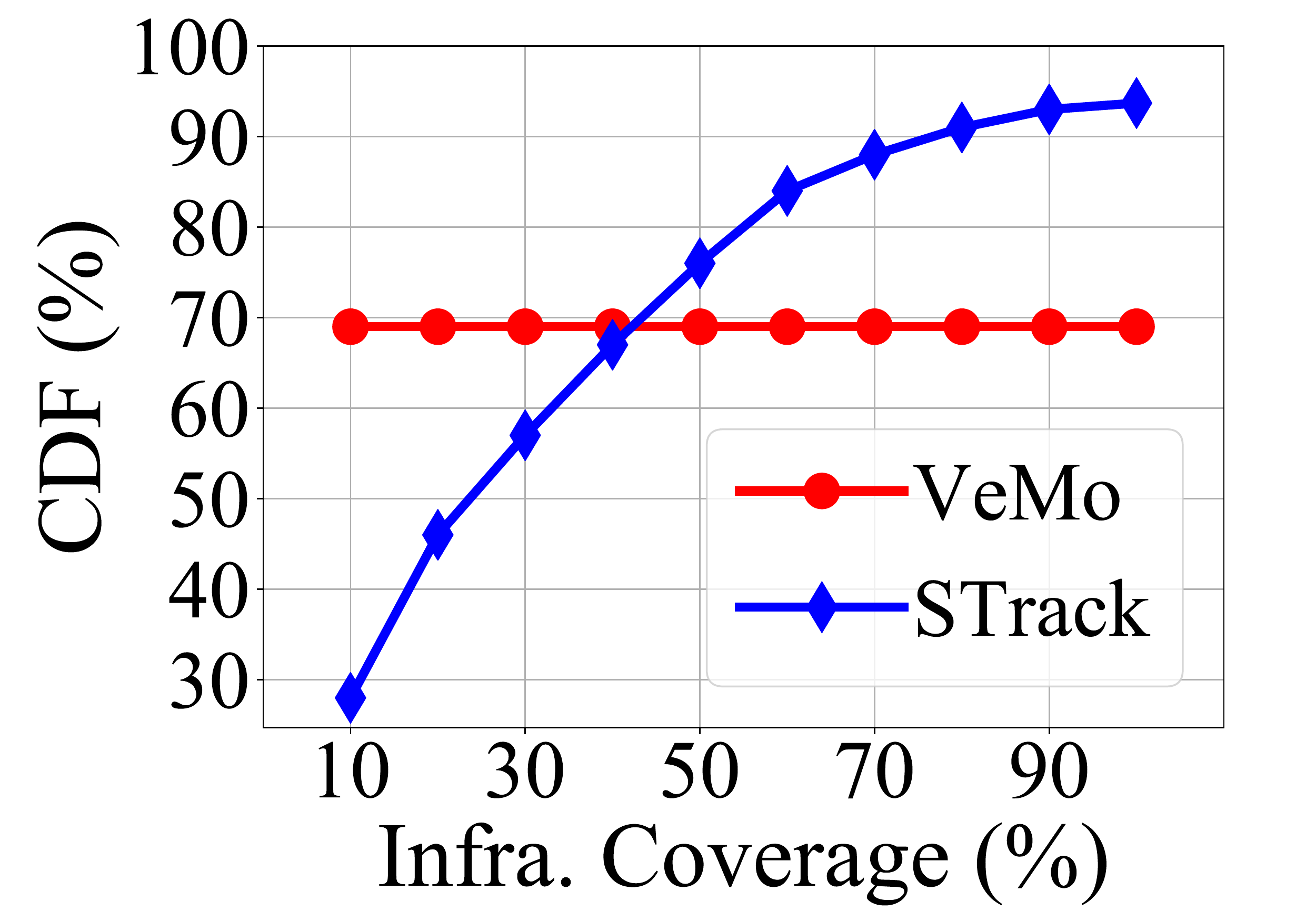}
        % \vspace*{-20pt}
        \caption{\% of infra.}
        \label{fig:evl_strack}
        % \hspace*{5pt}
    \end{minipage}
    \hspace*{-5pt}
    \begin{minipage}{0.5\linewidth} \centering
        \includegraphics[width=\linewidth, keepaspectratio=true]{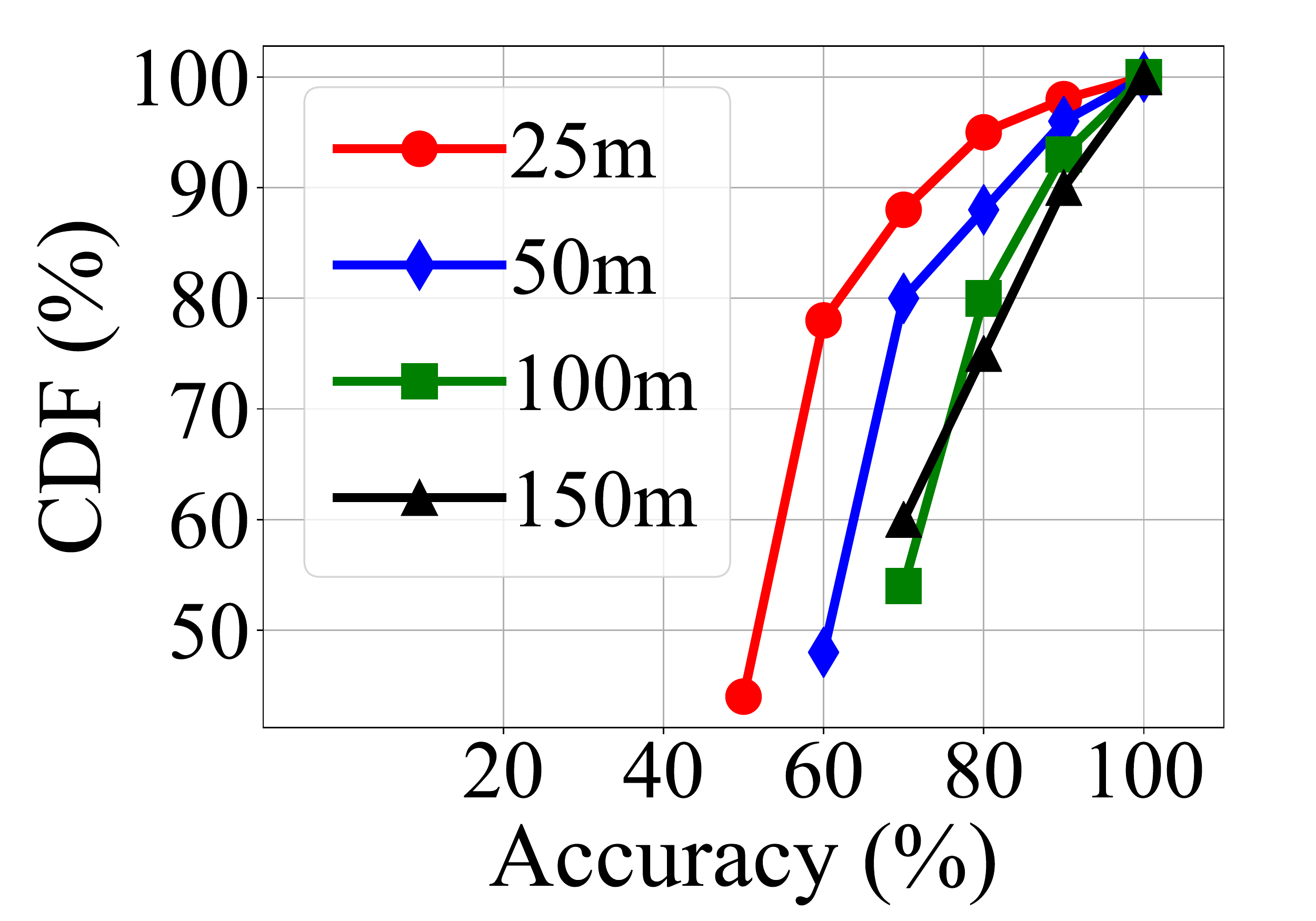}
        % \vspace*{-20pt}
        \caption{Accuracy threshold}
        \label{fig:thres_factors}
    \end{minipage}
    % \vspace*{-5pt}
\end{figure}

\noindent\textbf{(i) Impacts of Accuracy Threshold:} We choose accuracy threshold including 25, 50, 100, 150 meters to show how the accuracy changes in Fig.~\ref{fig:thres_factors}. The lower the line, the better the accuracy. We found higher thresholds lead to higher accuracy. 100-meter and 150-meter thresholds have closed accuracy while 25-meter and 100-meter thresholds have obvious lower accuracy.

\noindent\textbf{(ii) Impacts of Weather:}
We select one day with heavy rain and compare the result with that of a regular day. We surprisingly found the rain even increase the prediction accuracy. Since people tend to drive slowly in the heavy rain, the individual speed is reduced and there is a smaller range of speed variance on the way, which benefits the prediction accuracy.

\begin{figure}[htbp] \centering
    % \vspace*{-5pt}
    \begin{minipage}{0.5\linewidth} \centering
        \includegraphics[width=\linewidth, keepaspectratio=true]{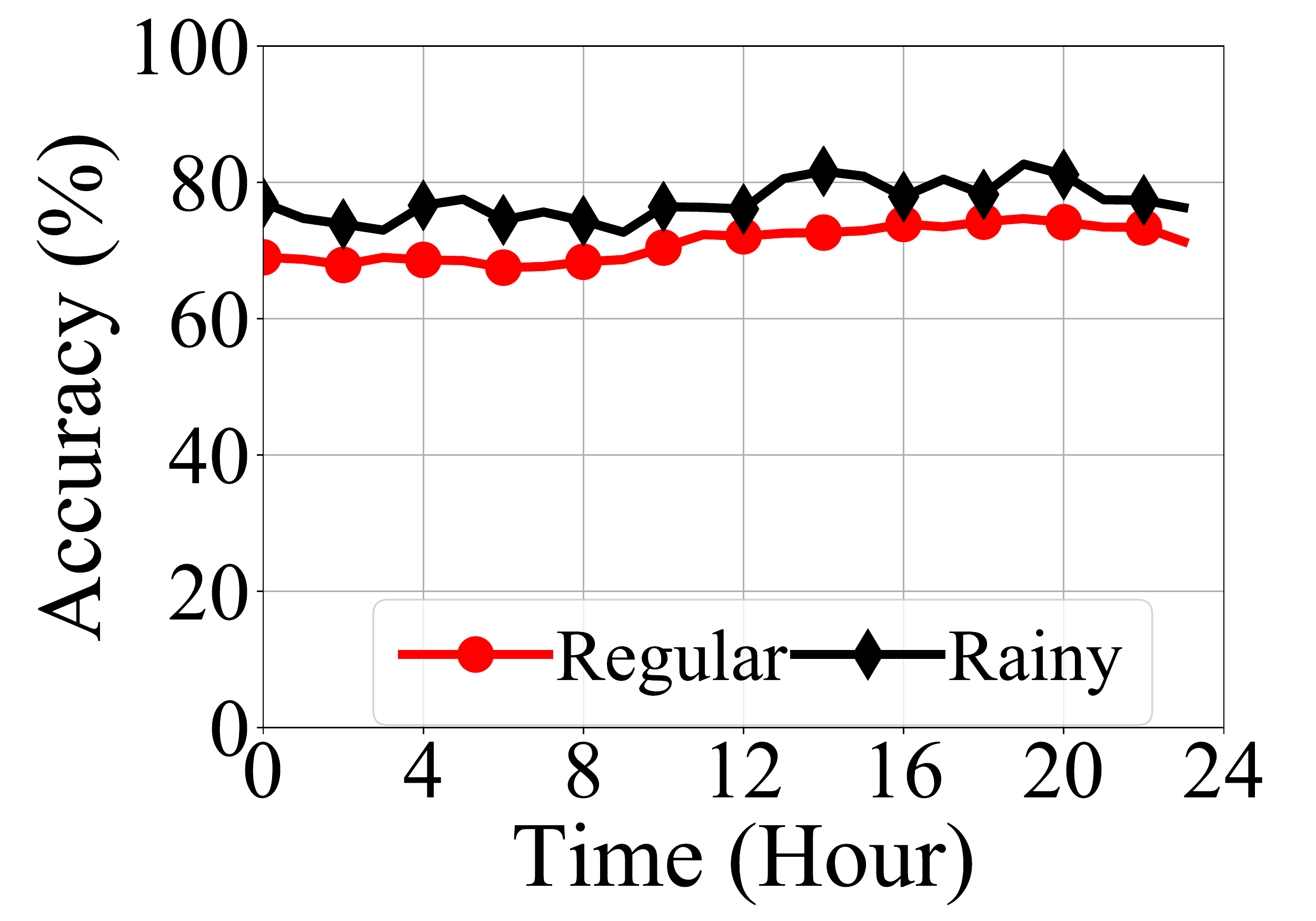}
        % \vspace*{-20pt}
        \caption{Weather Impact}
        \label{fig:weather_impact}
        % \hspace*{5pt}
    \end{minipage}
    \hspace*{-5pt}
    \begin{minipage}{0.5\linewidth} \centering
        \includegraphics[width=\linewidth, keepaspectratio=true]{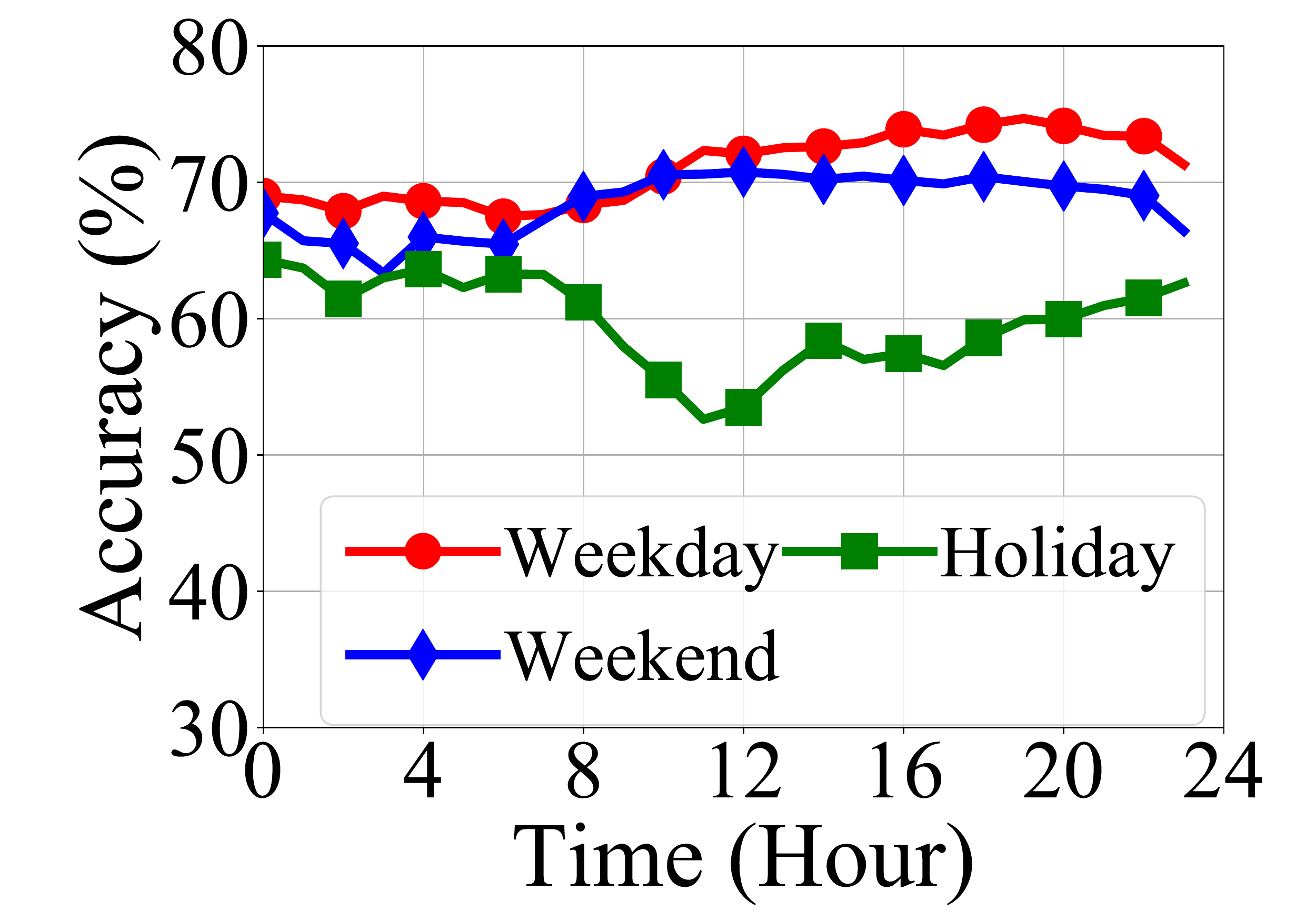}
        % \vspace*{-20pt}
        \caption{Time impact}
        \label{fig:time_factors}
    \end{minipage}
    % \vspace*{-5pt}
\end{figure}

\noindent\textbf{(iii) Impacts of Time Factors:} Fig~\ref{fig:time_factors} shows the performance of VeMo in weekday, weekend and holiday. The accuracy in weekday and weekend is similar. Moreover, the performance in the holiday is different than other days, especially during the morning. This is because the destination choices are less predictable on the holidays when people generally do not follow regular mobility patterns.

\begin{figure}[htbp] \centering
    % \vspace*{-5pt}
    \begin{minipage}{0.5\linewidth} \centering
        \includegraphics[width=\linewidth, keepaspectratio=true]{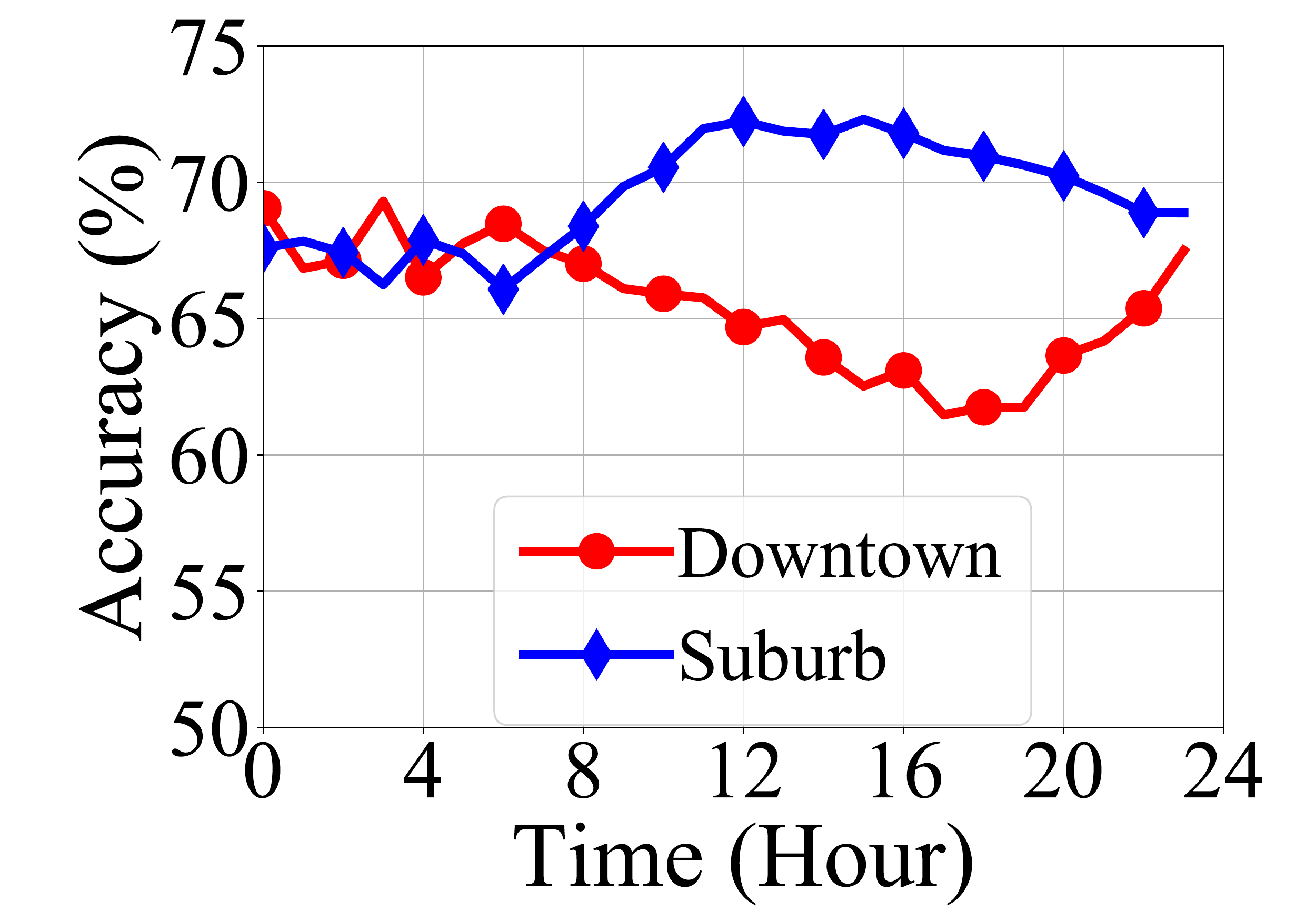}
        % \vspace*{-20pt}
        \caption{Spatial impact}
        \label{fig:spatial_factors}
    \end{minipage}
    \hspace*{-5pt}
    \begin{minipage}{0.5\linewidth} \centering
        \includegraphics[width=\linewidth, keepaspectratio=true]{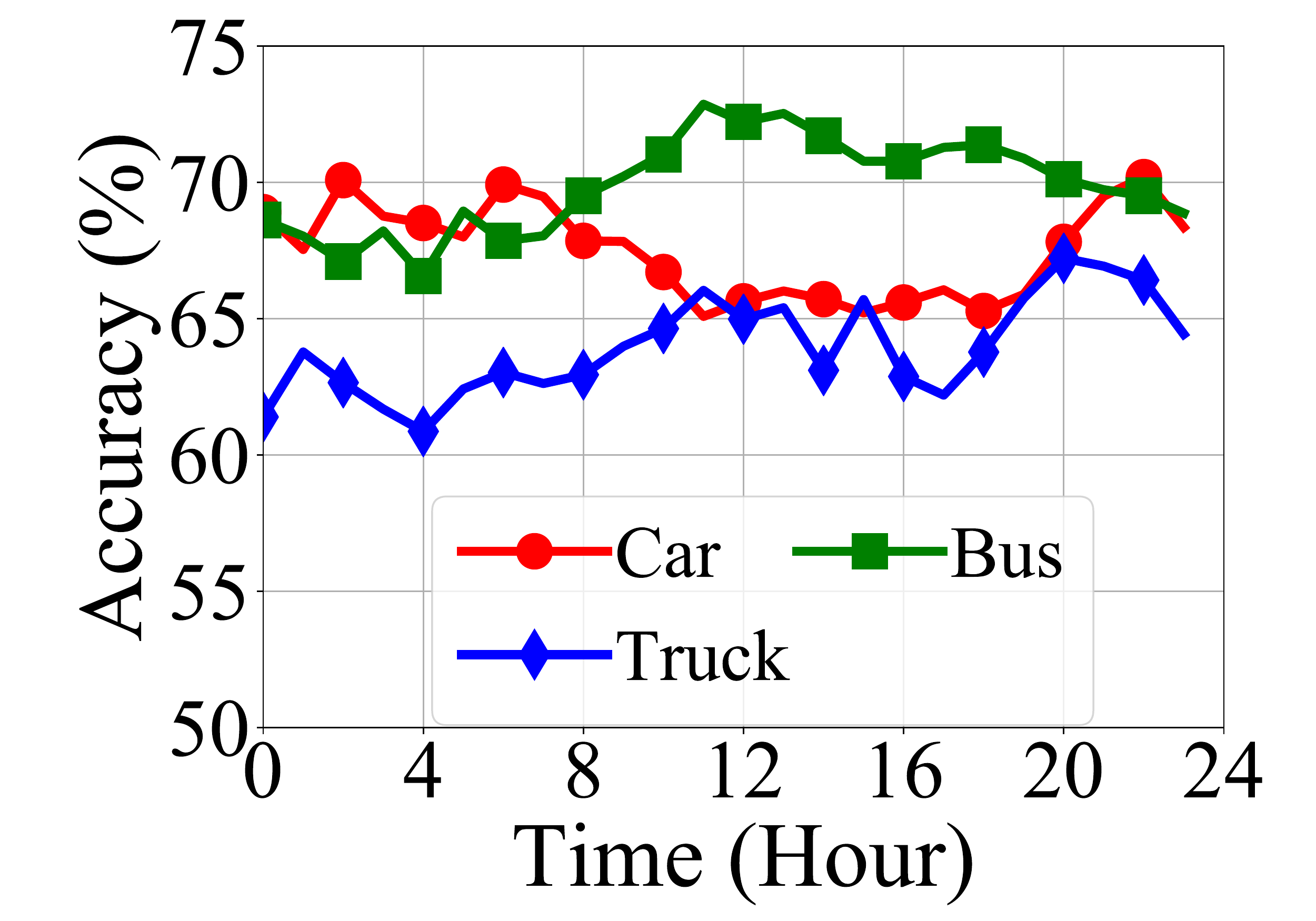}
        % \vspace*{-20pt}
        \caption{Type impact}
        \label{fig:vehicle_type_factors}
    \end{minipage}
    % \vspace*{-5pt}
\end{figure}

\noindent\textbf{(iv) Impacts of Spatial Factors:} We investigate the performance of VeMo in both downtown areas and suburb areas and show the result in Fig~\ref{fig:spatial_factors}. In the early morning, two areas have similar accuracy. During the daytime starting at 8 am, the performance in the downtown areas decreases. This is because the road structure is more complex in that areas, which makes the route prediction less accurate.

\noindent\textbf{(v) Impacts of Vehicle Types:}
Fig~\ref{fig:vehicle_type_factors} shows the performance of different types of vehicles. Trucks have the lowest accuracy because they generally have longer travel distances and irregular mobility patterns (e.g., one truck may travel between different areas for cargo services as long as there are demands of cargo transportation). Buses have higher accuracy because they have most regular mobility patterns compared to trucks and cars. Cars' accuracy decreases during the daytime because they generally travel inner cities, which is impacted by both traffic conditions and road structures.

\section{Discussions}
\label{sec:Dis}

\noindent\textbf{Lessons Learned:}
Based on our results in Guandong, we learned a few valuable lessons.
\begin{itemize}
    \item vehicle's mobility pattern in terms of destinations can be identified as three major groups, single-time travel vehicles, commuting vehicles and multi-destination vehicles;
    \item the overall distributions of both speed STDs cross vehicles and speeds on the road segment follow strong normality, which can be considered as constrains to infer the routes and speeds simultaneously;
    \item individual highway speeds vary based on driving behaviors, but is highly correlated with generic traffic speeds. The overall derivation of individual speeds follows a Gaussian-like distribution;
    \item Both individual and crowd level features (e.g., historical destinations, routes, vehicle types, driving experience, etc ) are helpful to predict vehicle locations along with contexts (e.g., time of day, day of week, weather).
\end{itemize}
% (1) vehicle's mobility pattern in terms of destinations can be identified as three major groups, single-time travel vehicles, commuting vehicles and multi-destination vehicles;
% (2) the overall distributions of both speed STDs cross vehicles and speeds on the road segment follow strong normality, which can be considered as constrains to infer the routes and speeds simultaneously;
% (2) for complicated highways, people may take several routes for the same origin and destination, which may not be the shortest or quickest route;
% (3) individual highway speeds vary based on driving behaviors, but is highly correlated with generic traffic speeds. The overall derivation of individual speeds follows a Gaussian-like distribution;
% (3) half of the vehicles have rather short distance (i.e., 22KM) and travel time (20 mins), with limited origin/destination patterns (i.e., 3) and 2 trips per week.
% (4) Both individual and crowd level features (e.g., historical destinations, routes, vehicle types, driving experience, etc ) are helpful to predict vehicle locations along with contexts (e.g., time of day, day of week, weather).

\noindent\textbf{Why ETC Data Only?}
In this work, we explore the possibility of using ETC data alone to predict real-time vehicle locations by solving some uncertainty issues, e.g., unknown routes.
This is because ETC systems can provide a full penetration rate transparently based on data already collected. 
Moreover, the ETC based toll system is universal and exist almost everywhere even in developing countries where satellite images or mobile infrastructure is not well penetrated.
If combined with other datasets even with small scale, 
e.g., GPS data from highway service vehicles or traffic camera data, we may be able to further improve our accuracy significantly.
But due to space limitation, we focus on our core contribution on ETC data.

\noindent\textbf{Data Collection and Privacy Protection:}
In this work, the ETC data we utilized are collected by an ETC company, which is a part of Guangdong Highway Administration Agency;
the GPS data we utilized are collected by an insurance company under drivers' agreement, which is a part of usage-based insurances for discounts.  In the ETC service agreement and highway usage agreement, people are notified that their data will be used to analyze traffic patterns and improve traffic condition.
Instead of tracking individual vehicles, our project is to understand and improve traffic patterns, which potentially benefits all ETC users.

\noindent\textbf{Real-world Applications:}
Our project is part of highway improvement initiative Guangdong Highway Administration Agency.
One key application of our vehicle location prediction is to address the traffic congestion on Guangdong highway during peak hours. Based on our results, we can estimate the number of vehicles on each edge of the highways by predicting the real-time locations of all vehicles, which can be utilized to design applications such as ramp meters~\cite{schmidt2015decentralised} and adaptive toll strategies~\cite{wired}.

\noindent\textbf{Rest Area Stop:}
Since the ETC data only give the station to station travel duration, 
it may contain time a person spent at rest areas, which cannot be directly obtained from the ETC transaction data. But based on our dataset, we found that 76\% of transactions have a duration less than 60 minutes, during which a person is unlikely to go to rest areas unless it is a part of a longer trip starting outside Guangdong Highway System.
Unfortunately, we cannot validate this assumption based on ETC data alone. 
However, with GPS data, we found that for the trips shorter than 60 minutes, only 8\% of vehicles went to rest areas.
It indicates rest area stops may not have significant impacts on our results.

\noindent\textbf{Limitations and Open Problems:} We discuss some limitations and open problems related to our system.
\begin{itemize}
    \item Each highway system has its unique geographic and demographic features, so data-driven insights and our evaluation results we have in Guangdong may not apply to other highways with very different features. However, we believe the techniques we develop to predict destinations, infer routes, and estimate speeds are generic and can be applied to other highway systems if their data are available.
    
    \item Our system work in a controlled environment, i.e., a highway system with both entering and existing records. Therefore the same technique may not be applied to local streets without toll booths to track every vehicle enter or leave a street. In this case, additional data, e.g., partial GPS, can be combined with our solution for prediction. However, we believe our solution can be generalized to stationary sensors that can capture the vehicle's passing (such as cameras, cellular tower, etc.).
    
    \item Even ETC systems only capture the vehicle twice, it still has the privacy issue of exposing locations. However, compared with GPS based solutions, it is a better/low-cost privacy-reserved approach since ETC data have already been collected as a mandatory process for billing; whereas other approaches need new devices or dedicate data collection process with potential continuous location collection.
    
    \item Our solution can help to detect the abnormal events when there are a certain number of vehicles (potentially spatial correlated vehicles, e.g., their real-time predicted locations are the same road segments) do not leave the highway after their travel duration on highways. But since in this work, we focused on the fundamental location prediction and did not try to explicitly handle the anomalies, which would be a good direction for our future work.
    
    \item Our solution replies on the historical data of vehicles to learn their driving behaviors. In the ETC system, we found there is only 9\% of the new vehicles without any historical data after 10-day data accumulation, which is very small number of vehicles. For those without historical data, we can only infer their behaviors according to majority behaviors (i.e., crowd features). It is still an interesting open problem.
\end{itemize}
% (2) We only predict locations of vehicles when they are on highways, but did not predict when they will enter highways due to space limitation. We argue that this prediction problem, though important, can be addressed by standard machine learning techniques.

\section{Related Work}
\label{sec:soa}

The most related work to this paper is a system called SharedEdge~\cite{yang2018sharededge} where ETC data are utilized to infer the generic traffic speed on each highway edge. 
However, VeMo is different in both the objective and methods. 
In particular, VeMo focuses on location prediction, whereas SharedEdge focuses on speed prediction. Even though VeMo also has a speed estimation but it focuses on individual speeds, whereas SharedEdge focuses on generic speed.  
Moreover, there a good body of literature for vehicle mobility modeling and predictions based on various sensing infrastructures~\cite{thiagarajan2011accurate}\cite{zhao2015vetrack}\cite{meng2017city}.
Shown in the Table~\ref{tb:soa}, we divide them into two major parts: GPS based approaches and None-GPS based approaches, where highlight our position with the full penetration.

\begin{table}[!htb]
\caption{Vehicular Mobility Survey}
\label{tb:soa}
\begin{tabular}{|c|c|c|c|}
\hline
\multicolumn{1}{|l|}{} & \begin{tabular}[c]{@{}c@{}}Aggregate  \end{tabular} & \multicolumn{2}{c|}{\begin{tabular}[c]{@{}c@{}}Individual \end{tabular}} \\ \hline
Mobile & ~\cite{zhang2017deep}~\cite{khezerlou2017traffic}~\cite{hoang2016fccf} & \multicolumn{2}{c|}{~\cite{ho2015pressure}~\cite{zhao2017greendrive}~\cite{thiagarajan2010cooperative}~\cite{aslam2012city}} \\ \hline
\multirow{2}{*}{Static} & \multirow{2}{*}{~\cite{meng2017city}~\cite{zhang2017fcn}~\cite{yang2018sharededge}} & \begin{tabular}[c]{@{}c@{}}Partial \\ Penetration\end{tabular} & \begin{tabular}[c]{@{}c@{}}Full \\ Penetration\end{tabular} \\ \cline{3-4} 
 &  & ~\cite{thiagarajan2009vtrack}~\cite{thiagarajan2011accurate}~\cite{chandrasekaran2011tracking} & Our work \\ \hline
\end{tabular}
% \vspace{-5pt}
\end{table}

\textbf{Static Infrastructure:}
Static infrastructures, e.g., traffic cameras~\cite{zhang2017fcn}, cell towers~\cite{thiagarajan2011accurate}, WiFi access point~\cite{thiagarajan2009vtrack}, are widely used for vehicle mobility modeling. Some communication related works are also studies based on the static infrastructure~\cite{balasubramanian2008interactive}~\cite{li2016mobileinsight}~\cite{bai2015tale}.
% In New York City, there are more than 643 closed circuit television cameras for real-time traffic monitoring~\cite{nyc_camera}.
% Cell towers signals are used~\cite{chandrasekaran2011tracking} to track the locations of smartphones, e.g., exploring WiFi devices to model the mobility~\cite{wu2017gain}. 
% Other signals such as millimeter wave radios~\cite{wei2015mtrack} are also applied for high-precision passive tracking.
% Some work also study the tracking of human~\cite{li2016practical}\cite{TMC16}.
However, one disadvantage of these approaches is either the lower coverage of infrastructures~\cite{cvp_media} or the low penetration of the apps that are used to interact with the infrastructure.
Without the installation of extra infrastructures for a full coverage, it is difficult to model and predict the mobility of all the vehicles. Most of these approaches require continuous movement detection, which is not always satisfied in the real world~\cite{li2015traffic}.
Compared with existing work, our approach makes use of the existing infrastructures to predict vehicle mobility without extra cost. All the vehicles entering the highways are detected, which does not require the installation of interaction apps.
The requirement of only single real-time observations, e.g., entrance to a highway, largely increases the feasibility of our approach in the real world.
Some approaches such as cell phone network may have potential to infer traffic condition with low cost. But normally the cellphone data are not available for highway administrators. They can only use the data collected by themselves. Further, the cell phone network cannot be narrowed down to vehicular mobility since the driver and passage cannot be distinguished from the cell phone data only, which may introduce extra bias. 

\textbf{Mobile Infrastructure:}
Mobile infrastructures, i.e., smartphones and onboard devices, are extensively studied to understand both individual and groups of vehicles. ~\cite{thiagarajan2009vtrack}~\cite{zhao2017greendrive}~\cite{ho2015pressure} use smartphones to track vehicles in real time. 
The inference on mobility is studied in details by smartphone data~\cite{thiagarajan2010cooperative}. ~\cite{aslam2012city} estimates the urban traffic using vehicular fleets with onboard devices.
Other works such as crowdsoucing information collection and energy issues can also benefit from the mobile infrastructures~\cite{chen2014privacy}~\cite{zhang2014vehicle}~\cite{mathur2010parknet}.
However, these approaches are either limited by low penetration rates of apps~\cite{statista} or focus on the aggregated level~\cite{aslam2012city}, and typically rise privacy issues of exposing vehicle GPS data~\cite{zhao2017greendrive}.

Vehicular mobility on the highways is also studied in the transportation community (i.e., the destination and speed prediction~\cite{chang1994recursive}~\cite{weng2014freeway}~\cite{wang2015short}~\cite{zhao2018travel}). However, previous works mainly focused on the aggregated traffic characteristics, such as origin-destination matrix or traffic speed on the road segments. Different from these works, our system aims at the mobility model of individual vehicles, which requires microscope analysis of the vehicle mobility pattern. Moreover, we utilize extra dataset as ground truth, which avoid the drawbacks of cross-validation in the previous works.

\textbf{Summary:}
Based on our discussion, most of the existing approaches are limited by additional deployment cost, low penetration rates, or requirement for privacy-prying GPS locations. 
In contrast, our approach makes use of the existing infrastructures with a full penetration rate to track individuals with only sparse location information, which makes our work significantly different from the existing approaches.

\section{Conclusion}
\label{sec:Con}
In this paper, we focus on vehicle location prediction on large-scale highway systems with sparse ETC data. 
In particular, we motivate and design a novel system called VeMo 
with three key technical components for the destination prediction, route inference, and speed estimation. 
More importantly, we implement and evaluate VeMo based on the large-scale data in the Guangdong highway network in China, utilizing an large-scale ETC system with 773 stations and a large-scale vehicle fleet with GPS data as ground truth.
We advance state-of-the-art vehicle mobility modeling approaches by some key lessons we learned.
We envision our results may benefit various applications including highway anomaly detection and risk assessment that we have been working with our partner.

\section*{Acknowledgments}
\label{sec: Acknowledgement}
The authors would like to thank anonymous reviewers for their valuable comments. This work is partially supported by the by Rutgers Research Council, Rutgers Global Center, China 973 Program (2015CB352400), National Natural Science Foundation of China (41401470).

\bibliographystyle{ACM-Reference-Format}
\bibliography{bibliography}

\end{document}